\documentclass[aps,pra,twocolumn,superscriptaddress,longbibliography,nofootinbib,floatfix]{revtex4-2}
\usepackage[T1]{fontenc}
\usepackage[utf8]{inputenc}
\usepackage{graphicx}
\usepackage{amsmath,amssymb,bm,mathtools}
\usepackage{xcolor}
\usepackage[hidelinks]{hyperref}
\usepackage{microtype}
\usepackage{booktabs}
\hypersetup{
  colorlinks = true,
  linkcolor  = blue,
  citecolor  = blue,
  urlcolor   = blue,
}
\newcommand{\Tr}{\operatorname{Tr}}
\newcommand{\diag}{\operatorname{diag}}
\newcommand{\RI}{\mathrm{RI}}
\newcommand{\EC}{\mathrm{EC}}
\newcommand{\fp}{\mathrm{fp}}
\newcommand{\CSE}{\mathcal{C}_{SE}}
\newcommand{\Sc}{\mathcal{S}_{c}}
\newcommand{\JQ}{J_Q}
\newcommand{\Pstar}{P_*}
\newcommand{\CSEstar}{\mathcal{C}_{SE}^{*}}
\newcommand{\Scstar}{\mathcal{S}_{c}^{*}}
\newcommand{\Qsupstar}{Q_{\mathrm{sup}}^{*}}
\newcommand{\JQstar}{J_Q^{*}}
\newcommand{\Sigmastar}{\Sigma_{\mathrm{reset}}^{*}}
\newcommand{\sigmastar}{\sigma^{*}}
\newcommand{\dd}{\mathrm{d}}
\newcommand{\ii}{\mathrm{i}}
\newcommand{\e}{\mathrm{e}}

\begin{document}

\title{Thermodynamics of Coherence-Selective Quantum Reset Protocols}

\author{Jishad Kumar}
\affiliation{MSP Group, Department of Applied Physics, Aalto University, P.O. Box 15600, FI-00076 Aalto, Espoo, Finland}
\author{Achilleas Lazarides}
\affiliation{Interdisciplinary Centre for Mathematical Modelling and Department of Mathematical Sciences, Loughborough University, Loughborough, Leicestershire LE11 3TU, United Kingdom}
\author{Tapio Ala-Nissila}
\affiliation{MSP Group, Department of Applied Physics, Aalto University, P.O. Box 15600, FI-00076 Aalto, Espoo, Finland}
\affiliation{Interdisciplinary Centre for Mathematical Modelling and Department of Mathematical Sciences, Loughborough University, Loughborough, Leicestershire LE11 3TU, United Kingdom}

\date{\today}

\begin{abstract}
We develop an exact theory of coherence-selective stroboscopic resetting for quadratic open quantum systems within the single-particle density-matrix formalism. We focus on the survival of coherences and the associated thermodynamic cost at the stroboscopic fixed point. To this end, we introduce a one-parameter family of reset channels that continuously interpolates between complete coherence erasure and complete coherence preservation. This unifies the reset-map description, the repeated-interaction and evolving-correlation endpoint channels, and the thermodynamic cost of environmental reinitialization. For a single fermionic level coupled to a structured semi-infinite tight-binding bath, we derive the exact affine stroboscopic map, solve for its unique fixed point, and compute the retained coherence spectrum, the post-reset occupation, and the reset heat current. We find that retained coherence increases monotonically with the retention parameter, whereas the reset heat current is generically nonmonotonic and is maximized at an intermediate operating point. Thus the protocol that stores the most coherence is not the one that dissipates the most heat. Exact operating diagrams further show that coherence-optimal and coherence-per-cost-optimal protocols are both driven toward the coherence-preserving endpoint, while the heat-optimal protocol depends strongly on the reset interval. We also show that this coherence-cost geometry survives at nonzero chemical potential as a filling-biased deformation of the same fixed-point tradeoff, rather than as an independent particle-current optimization problem. These results establish coherence-selective resetting as a distinct control principle for structured-bath open quantum systems and provide an exactly solvable benchmark for memory engineering and thermodynamic optimization under repeated environmental reinitialization.
\end{abstract}

\maketitle

\section{Introduction}

A powerful formalism to describe open quantum systems is based on the idea of repeated interactions, where the unitary dynamics of a closed system is modified by a non-unitary operation due to interaction with an environment or heat bath. Such repeated interventions can profoundly reshape quantum evolution. In its traditional formulation, the quantum Zeno effect describes the suppression of state decay by sufficiently frequent measurements, while the anti-Zeno effect describes the opposite acceleration of decay under appropriately timed interventions~\cite{Misra1977,Schulman1998,Fischer2001,Home1997,Kofman2000Nature,Kofman2000Fortschr,Ruseckas2004,Facchi2001,Venugopalan2012,Itano2009}. Over the years this basic idea has been generalized well beyond ideal projective observations to include continuous monitoring, dynamical decoupling, dephasing-assisted control, and open-system interventions that repeatedly disturb the system-environment coupling structure~\cite{Itano1990,Streed2006,Facchi2008,Facchi2002,Zhou2017,Shaji2004,Alter1997,Frerichs1991,Gagen1992,Beige1996,Chandrashekar2010,Li2013,Maniscalco2006}. In parallel, collision models and repeated-interaction constructions have become a standard microscopic language for open-system control, nonequilibrium steady states, memory effects, and quantum thermodynamics~\cite{Becker2022,Ciccarello2022,Cusumano2022,Giovannetti2012,Scarani2002,Strasberg2017,Strasberg2019,Ciccarello2013,Kretschmer2016,Lorenzo2017,Giovannetti2005,Burgarth2007,Cattaneo2021,Cakmak2019}.

For quadratic fermionic and bosonic systems, these ideas become transparent because the dynamics closes exactly at the level of the single-particle density matrix (SPDM). In this setting, one does not need to write a Markovian master equation or invoke an auxiliary weak-coupling approximation: the entire stroboscopic dynamics can be written directly as an affine map acting on the kept SPDM entries. This exact SPDM reset-map framework was formulated in Ref.~\onlinecite{Vieira2020}, where two important limiting protocols were proposed. In the repeated-interaction (RI) protocol, the environment is reset after every cycle and all system-environment coherences are erased. In the evolving-correlations (EC) protocol, the environment block is reset but the system-environment coherences are retained. Our recent study of Zeno and anti-Zeno physics in this framework revealed that {\it coherence erasure vs. coherence retention} is the organizing principle behind the frequent-reset asymptotics and behind the thermodynamic cost of enforcing the reset~\cite{Kumar2026Zeno}. Related repeated-interaction studies have emphasized the emergence of stationary states and the energetic cost of reinitialization~\cite{Prositto2025,Strasberg2017,Strasberg2019,Allahverdyan2001,Guryanova2020,Deffner2016,Jacobs2012,Abdelkhalek2016,Landi2021}.

The present work is motivated by a simple but, to our knowledge, unexplored aspect of the reset dynamics. The $\RI$ and $\EC$ protocols differ only in the way they treat the system-environment coherence block. From the viewpoint of exact SPDM dynamics, a natural generalization is to reset the environment exactly as before but to preserve a finite and tunable fraction of the SPDM coherence blocks. This raises new interesting questions about how much of the system-environment memory is retained in the exact stroboscopic fixed point, how that memory is spectrally distributed, and how the answer is constrained by the thermodynamic cost of the reset channel.

To this end, in the present work we introduce a one-parameter family of \emph{coherence-selective} reset channels labeled by $\eta\in[0,1]$. The limit $\eta=0$ corresponds to the $\RI$ rule of complete coherence erasure, while $\eta=1$ reproduces the $\EC$ rule of complete coherence retention. Intermediate values describe a partial reset of the coherences. Because the Hamiltonian remains quadratic, the resulting stroboscopic dynamics is still exactly soluble at the SPDM level. We analyze the simplest nontrivial instance in full detail: a single fermionic level coupled to a semi-infinite tight-binding chain. This model is a standard structured bath benchmark with a bounded spectral density and sharp band edges~\cite{Harrison1989,Grosso2014,Colombo2005,Pettifor1989,Hoekstra1988,Sedlmayr2013,Economou2006}.

The main result is an exact fixed-point coherence-cost tradeoff for the reset channel itself. The retained coherence grows monotonically with $\eta$, whereas the exact reset heat current is generically nonmonotonic in $\eta$. The consequence is operationally sharp: the reset channel that stores the most coherence is not the reset channel that dissipates the most heat. Moreover, the channels that maximize retained coherence and coherence-per-cost ratio are both driven toward the coherence-preserving endpoint, whereas the heat-optimal channel sits at an intermediate value of $\eta$ that drifts markedly with the stroboscopic interval. The exact operating diagrams therefore separate three optimization tasks that coincide neither dynamically nor thermodynamically. We also extend the analysis beyond the particle-hole-symmetric benchmark by allowing a nonzero chemical potential $\mu$ in the reset bath state. This introduces a controlled filling bias that reshapes the exact fixed-point coherence and heat-current landscapes through the self-consistent occupation mismatch entering the reset map. As we show below, the $\mu\neq0$ case deforms the same
coherence-cost geometry already present at $\mu=0$, rather than generating an independent fixed-point particle-transport tradeoff.

The paper is organized as follows. In Sec.~\ref{sec:model} we define the quadratic model and the $\eta$-dependent reset rule. In Sec.~\ref{sec:fixed-point} we derive the exact one-cycle affine map and the fixed-point equations for a single level coupled to a bath. Section~\ref{sec:tb} specializes the bath to a semi-infinite tight-binding chain and defines the exact retained-coherence and thermodynamic observables. Section~\ref{sec:results} presents the exact numerical results and the operating diagrams. Section~\ref{sec:mu_extension} discusses the nonzero-chemical-potential case as a filling-biased deformation of the same fixed-point coherence-cost geometry. Section~\ref{sec:discussion} discusses the physical meaning of the tradeoff and possible extensions and conclusions. Detailed derivations are collected in Appendices~\ref{app:map}-\ref{app:numerics}.

\section{Quadratic model, reset maps, and thermodynamic cost}
\label{sec:model}

\subsection{Quadratic dynamics at the SPDM level}

We consider a number-conserving quadratic fermionic Hamiltonian
\begin{equation}
\hat H = \sum_{\alpha,\beta} \hat a_{\alpha}^{\dagger} M_{\alpha\beta} \hat a_{\beta},
\label{eq:Hquad}
\end{equation}
where $M=M^{\dagger}$ is the one-particle Hamiltonian matrix. The corresponding single-particle density matrix (SPDM) is
\begin{equation}
\rho_{\alpha\beta}(t)=\langle \hat a^\dagger_{\beta}\hat a_{\alpha}\rangle_t .
\label{eq:spdmdef}
\end{equation}
Because the Hamiltonian is quadratic, the SPDM evolves exactly under the one-particle propagator
\begin{equation}
U(t)=e^{-\ii M t}
\label{eq:Ut}
\end{equation}
as
\begin{equation}
\rho(t)=U(t)\,\rho(0)\,U^\dagger(t).
\label{eq:spdmevol}
\end{equation}
Thus the entire stroboscopic dynamics can be formulated directly in terms of finite-dimensional matrix algebra, with no auxiliary closure approximation and no weak-coupling reduction. This exact SPDM language is the natural common framework for the earlier repeated-interaction and evolving-correlation protocols and for the coherence-selective channel introduced below.

We partition the single-particle Hilbert space into a system sector $S$ and an environment sector $E$ and write
\begin{equation}
\rho=\begin{pmatrix}
\rho_{SS} & \rho_{SE}\\
\rho_{ES} & \rho_{EE}
\end{pmatrix};
\quad
M=\begin{pmatrix}
M_{SS} & M_{SE}\\
M_{ES} & M_{EE}
\end{pmatrix}.
\label{eq:blockdecomp}
\end{equation}
A stroboscopic cycle has duration $\tau$. If $t_n=n\tau$ denotes the $n$th reset time, then $\rho_n\equiv \rho(t_n^+)$ is the SPDM immediately after the $n$th reset, and the exact pre-reset SPDM at the end of the next unitary segment is
\begin{equation}
\rho_n^- \equiv \rho(t_{n+1}^-)=U(\tau)\rho_n U^{\dagger}(\tau).
\label{eq:rhonminus}
\end{equation}
At this stage it is useful to clarify the meaning of the stroboscopic dynamics. The present problem is not a single-fermion evolution problem but a quadratic many-fermion Gaussian problem, fully characterized by its SPDM. Accordingly, the quantity \(N_{\mathrm{tot}}(t)=\Tr\,\rho(t)\) is the total \emph{expected} particle number in the system-plus-environment sector, not a fixed value equal to one. Because the Hamiltonian is number conserving, the unitary segment of each cycle conserves $N_{\mathrm{tot}}$ exactly. The reset step, however, is nonunitary: it restores the bath block to the reference SPDM and therefore need not conserve the total particle number of the system-plus-environment subsystem over a full cycle. The natural initial condition for the stroboscopic dynamics is an arbitrary physically admissible \emph{post-reset} SPDM $\rho_0\equiv\rho(t_0^+)$. The main results of this work, however, are not transient-state results; they are fixed-point results. That is, we study the post-reset state that is reproduced after one full unitary-plus-reset cycle and the associated pre-reset state obtained from it after one unitary segment. All starred quantities introduced in the thermodynamics study, later, are evaluated at this exact stroboscopic fixed point. For the reset protocol considered here, the fixed-point particle current vanishes, so the nonzero-$\mu$ extension deforms the coherence-cost geometry through the self-consistent occupation mismatch rather than through an independent steady particle-transport channel.

\subsection{General stroboscopic reset map and the RI/EC endpoints}
\label{subsec:stroboscopic_map}

At the SPDM level a reset operation is a linear map that overwrites a specified set of matrix elements with prescribed reference values. Let $R\subset \{1,\dots,N\}^2$ be the set of ordered index pairs $(\alpha,\beta)$ that are reset after each unitary step, and let ${\cal K}=R^{\rm c}$ denote the complementary set of entries that are kept. Defining a vector of kept SPDM entries,
\begin{equation}
V_i[n] \equiv \rho_{\alpha_i\beta_i}(t_n),
\end{equation}
where $i\mapsto (\alpha_i,\beta_i)\in {\cal K}$ is a fixed enumeration, the exact unitary evolution over one cycle gives
\begin{equation}
\rho_{\alpha\beta}(t^-_{n+1})
=
\sum_{\alpha',\beta'}
U_{\alpha\alpha'}(\tau)\,
U^{*}_{\beta\beta'}(\tau)\,
\rho_{\alpha'\beta'}(t_n).
\label{eq:component_update}
\end{equation}
For kept entries we retain the unitary-updated value, while reset entries are replaced by their reference values $\rho_{\alpha\beta}^{(0)}$. One then obtains the exact affine map
\begin{equation}
V[n+1] = D(\tau)V[n] + C(\tau),
\label{eq:stroboscopic_map}
\end{equation}
with matrix elements
\begin{align}
D_{ij}(\tau) &= U_{\alpha_i\alpha_j}(\tau)\,
U^{*}_{\beta_i\beta_j}(\tau);\\
C_i(\tau)
&=
\sum_{(\alpha',\beta')\in R}
U_{\alpha_i\alpha'}(\tau)\,
U^{*}_{\beta_i\beta'}(\tau)\,
\rho^{(0)}_{\alpha'\beta'} .
\end{align}
Equation~\eqref{eq:stroboscopic_map} is the central exact SPDM reset-map framework used throughout the paper. The model dependence enters through the quadratic propagator $U(\tau)=e^{-\ii M\tau}$ and the choice of which entries are overwritten.

Two limiting reset rules are especially important. In the $\RI$ protocol, the environment block is reset and the system-environment coherence blocks are fully erased after every cycle as
\begin{equation}
\rho_{EE}^+=\rho_{EE}^{(0)};\qquad \rho_{SE}^+=\rho_{ES}^+=0.
\end{equation}
In the evolving-correlation protocol, denoted $\EC$, the environment block is reset but the coherence blocks are kept,
\begin{equation}
\rho_{EE}^+=\rho_{EE}^{(0)},\qquad \rho_{SE}^+=\rho_{SE}^-,\qquad \rho_{ES}^+=\rho_{ES}^-.
\end{equation}
Our earlier work showed that this distinction is decisive for the short-time dynamical asymptotics and for the thermodynamic cost of resetting: erasing versus preserving $S-E$ coherence is the organizing principle behind the RI/EC dichotomy~\cite{Kumar2026Zeno}.

\subsection{Coherence-selective reset rule}

The reset channel studied here leaves the system block unchanged, restores the environment block to a fixed reference SPDM, and multiplies the system-environment coherence blocks by a control parameter $\eta\in[0,1]$:
\begin{subequations}
\begin{align}
\rho_{SS}(t_{n+1}^+) &= \rho_{SS}(t_{n+1}^-);
\label{eq:etareseta}\\
\rho_{EE}(t_{n+1}^+) &= \rho_{EE}^{(0)};
\label{eq:etaresetb}\\
\rho_{SE}(t_{n+1}^+) &= \eta\,\rho_{SE}(t_{n+1}^-);
\label{eq:etaresetc}\\
\rho_{ES}(t_{n+1}^+) &= \eta\,\rho_{ES}(t_{n+1}^-).
\label{eq:etaresetd}
\end{align}
\label{eq:etareset}
\end{subequations}
The reference environment SPDM is denoted by $C_0\equiv \rho_{EE}^{(0)}$. In thermal equilibrium and in the eigenbasis of $M_{EE}$ one has
\begin{equation}
C_0 = \diag(n_1,n_2,\dots);
\quad
n_k = \frac{1}{\e^{\beta(\omega_k-\mu)}+1}.
\label{eq:C0}
\end{equation}
The endpoint $\eta=0$ reproduces $\RI$, since the coherence blocks are fully erased after each step. The other endpoint $\eta=1$ reproduces $\EC$, since the coherence blocks are kept while the environment occupations are reset. Intermediate values of $\eta$ continuously interpolate between these two exact protocols. The interpolation is therefore not heuristic: it is defined directly at the SPDM level by the same reset-map logic as the RI/EC endpoints.

\subsection{Thermodynamic cost of a reset}
\label{subsec:thermo_cost}

The reset step is nonunitary and therefore carries an energetic and entropic cost. Following the super-environment interpretation used in our earlier resetting work, we attribute that cost to an external agent that restores the bath block from its pre-reset value to the reference SPDM $C_0$~\cite{Jacobs2012,Abdelkhalek2016,Deffner2016}. The advantage of the SPDM formulation is that this cost can be expressed exactly and directly at the single-particle level, without invoking weak-coupling master equations or additional coarse-graining approximations. To be precise, we write the quadratic Hamiltonian as
\begin{equation}
\hat H = \hat H_S + \hat H_E + \hat H_{\mathrm{int}},
\end{equation}
with $\hat H_S$ and $\hat H_E$ quadratic in system and environment operators, respectively, and $\hat H_{\mathrm{int}}$ containing the quadratic couplings between $S$ and $E$. In terms of the single-particle Hamiltonian matrix $M$ this corresponds to the block splitting
\begin{equation}
    M=
\begin{pmatrix}
M_{SS} & M_{SE}\\
M_{ES} & M_{EE}
\end{pmatrix}.
\end{equation}
For any Gaussian state, the average environment energy at time $t$ is
\begin{equation}
E_E(t)\equiv \mathrm{Tr}_E\big(\hat H_E\hat\rho(t)\big)=\mathrm{Tr}_E\!\left(M_{EE}\rho_{EE}(t)\right),
\label{eq:EEdef}
\end{equation}
where the trace on the right-hand side is taken over environment indices only and we used
$\langle \hat a^\dagger_\beta \hat a_\alpha\rangle=\rho_{\alpha\beta}$.
During the unitary step from $t_n$ to $t^-_{n+1}$ the total energy is conserved,
\begin{equation}
\mathrm{Tr}\big(\hat H\,\hat\rho(t^-_{n+1})\big)=\mathrm{Tr}\big(\hat H\,\hat\rho(t_n)\big).
\end{equation}
The reset step from $t^-_{n+1}$ to $t_{n+1}$ is non-unitary and changes the environment energy from $E_E(t^-_{n+1})$ to $E_E(t_{n+1})$. We interpret this change as heat exchanged with a super-environment enforcing the reference state $\rho^{(0)}_{EE}$ \cite{Jacobs2012,Abdelkhalek2016,Deffner2016}.

We define the heat dumped into the super-environment during the $n$th reset as
\begin{align}
Q^{(n)}_{\mathrm{sup}}
&\equiv
-\Big\langle E_E(t_{n+1})-E_E(t^-_{n+1})\Big\rangle \nonumber\\
&=
\mathrm{Tr}_E\!\left[
M_{EE}\big(\rho_{EE}(t^-_{n+1})-\rho_{EE}(t_{n+1})\big)
\right].
\label{eq:qsupp}
\end{align}
A positive $Q^{(n)}_{\mathrm{sup}}$ means that energy has been removed from the $S{+}E$ system and dumped into the super-environment. In both RI and EC, by construction, the environment block after the reset is set to the same
reference SPDM,
\begin{equation}
\rho_{EE}(t_{n+1})=\rho^{(0)}_{EE} \equiv C_0,
\end{equation}
and thus Eq.~(\ref{eq:qsupp}) reduces to the compact expression
\begin{equation}
Q^{(n)}_{\mathrm{sup}}
=
\mathrm{Tr}_E\!\left[
M_{EE}\big(\rho_{EE}(t^-_{n+1})-\rho^{(0)}_{EE}\big)
\right].
\label{eq:qsupp_new}
\end{equation}
Note that in both reset protocols the system block is not changed by the reset, $\rho_{SS}(t_{n+1})=\rho_{SS}(t^-_{n+1})$, hence the reset-induced energy change is purely environmental. Equation~\eqref{eq:qsupp_new} is exact and protocol independent: the protocol dependence enters only through the pre-reset bath block $\rho_{EE}(t^-_{n+1})$, which is determined by the stroboscopic fixed point. 

To discuss the cycle-stationary thermodynamics of the reset channel, we will use a superscript $*$ to denote the \emph{stroboscopic fixed point} (or cycle-stationary state) of the one-cycle reset map. Concretely, a fixed point is a post-reset state that is reproduced after one full unitary-plus-reset cycle. Given such a fixed-point post-reset state, the associated pre-reset state is obtained by propagating it through one unitary segment of duration $\tau$ before the next reset is applied. We denote the corresponding pre-reset bath block by
\begin{equation}
C_{\mathrm{fp}}^{-}(\tau,\eta)\equiv \rho_{EE}(t_{n+1}^{-})\big|_{\mathrm{fp}} .
\end{equation}

The transient dynamics may start from any physically admissible post-reset SPDM, but all thermodynamic quantities plotted in this work are evaluated only after the stroboscopic fixed point has been reached. All starred thermodynamic quantities below are evaluated from this exact cycle-stationary state. When the reset map has reached its cycle-stationary state, the corresponding fixed-point heat per reset is
\begin{equation}
\Qsupstar(\tau,\eta)=\Tr_E\!\left[M_{EE}\big(C_{\fp}^{-}(\tau,\eta)-C_0\big)\right],
\label{eq:Qsupstar_sec2}
\end{equation}
and the associated heat current is
\begin{equation}
\JQstar(\tau,\eta)=\frac{\Qsupstar(\tau,\eta)}{\tau}.
\label{eq:JQstar_sec2}
\end{equation}
Here $C_{\mathrm{fp}}^{-}(\tau,\eta)\equiv \rho_{EE}(t_{n+1}^{-})\big|_{\mathrm{fp}}$. Thus the star denotes a quantity evaluated at the exact stroboscopic fixed point of the reset channel, while the superscript $(n)$ refers to a single reset event before the fixed-point limit is taken.

For a Gibbs reference environment,
\begin{equation}
\hat\rho_E^{(0)}
=
\frac{e^{-\beta(\hat H_E-\mu \hat N_E)}}{Z_E};
\quad
Z_E=\Tr\,e^{-\beta(\hat H_E-\mu \hat N_E)} .
\label{eq:gibbs}
\end{equation}
The corresponding entropy production of the reset operation is quantified by the quantum relative entropy \cite{Spohn1978EntropyProduction,DeffnerLutz2011EntropyProduction}
\begin{equation}
\Sigma_{\mathrm{reset}}^{(n)} = D\!\left(\hat\rho_E(t^-_{n+1})\Vert \hat\rho_E^{(0)}\right).
\label{eq:sigma_one}
\end{equation}
The entropy-production rate associated with the exact fixed point is 
\begin{equation}
  \sigmastar(\tau,\eta)=\Sigmastar(\tau,\eta)/\tau  ,
\end{equation}
while for a single reset event one may also define the transient rate $\Sigma_{\mathrm{reset}}^{(n)}/\tau$. For Gaussian fermionic bath states, Eqs.~\eqref{eq:qsupp_new} and \eqref{eq:sigma_one} can be written directly in terms of the environment SPDMs $C=\rho_{EE}(t^-_{n+1})$ and $C_0=\rho_{EE}^{(0)}$ \cite{HacklBianchi2021GaussianKahler,Higham2008FunctionsOfMatrices,PtaszynskiEsposito2023GaussianEntropyProduction,PeschelEisler2009FreeLatticeRDM}:
\begin{equation}
S(C)=-\Tr_E\big[C\ln C + (I-C)\ln(I-C)\big];
\label{eq:SCmain}
\end{equation}
\begin{align}
D(C\Vert C_0)&=\Tr_E\Big[C(\ln C-\ln C_0)\nonumber\\&+(I-C)\big(\ln(I-C)-\ln(I-C_0)\big)\Big].
\label{eq:DCmain}
\end{align}
Moreover, for the Gibbs reference, one has the exact decomposition
\begin{align}
\Sigma_{\mathrm{reset}}^{(n)}
&=
\beta\!\left(Q_{\mathrm{sup}}^{(n)}-\mu\,\Delta N_E^{(n)}\right)
-
\bigl[S(C)-S(C_0)\bigr];
\\
\Delta N_E^{(n)}&=\Tr_E(C-C_0)\nonumber.
\label{eq:rel_entropy_newdef}
\end{align}
These formulae are the thermodynamic starting point for the coherence-cost tradeoff developed below. The novelty of the present work is that the retained coherence block, controlled by $\eta$, feeds directly into the pre-reset bath block and therefore into both the heat current and the entropy production. 

At the fixed point the reset entropy per cycle and the corresponding entropy-production rate are
\begin{equation}
\Sigmastar(\tau,\eta)=D\!\left(C_{\fp}^{-}(\tau,\eta)\Vert C_0\right);
\label{eq:Sigmastar_sec2}
\end{equation}
\begin{equation}
\sigmastar(\tau,\eta)=\frac{\Sigmastar(\tau,\eta)}{\tau}.
\label{eq:sigmastar_sec2}
\end{equation}
The same definitions apply to RI, EC, and the interpolating coherence-selective reset channel.
What changes from one protocol to another is not the thermodynamic bookkeeping itself, but the exact stroboscopic state that produces the pre-reset bath block $C_{\mathrm{fp}}^{-}(\tau,\eta)$. In the RI endpoint, the reset erases the system-environment coherence block and the resulting pre-reset bath state generally differs greatly from the reference state. In the EC endpoint, by contrast, the reset preserves the propagated system-environment coherence block and modifies only the bath occupations. This difference is already visible at the level of the endpoint thermodynamics.

\section{Exact one-cycle map and fixed point for a single level: zero chemical potential case}
\label{sec:fixed-point}

\subsection{One system level coupled to a bath}

We now specialize to one system level labeled by $0$, coupled to bath modes $k\in E$:
\begin{align}
M = \omega_0 |0\rangle\langle 0| &+ \sum_k \omega_k |k\rangle\langle k| \nonumber\\&+ \sum_k\big(g_k |0\rangle\langle k| + g_k^{*}|k\rangle\langle 0|\big).
\label{eq:onelevelM}
\end{align}
The system block is then a single number, namely the occupation
\begin{equation}
P_n \equiv \rho_{00}(t_n^+).
\label{eq:Pndef}
\end{equation}
The post-reset coherence blocks are written as a row vector and its Hermitian conjugate,
\begin{equation}
\mathbf{x}_n \equiv \rho_{0E}(t_n^+);
\qquad
\mathbf{y}_n \equiv \rho_{E0}(t_n^+)=\mathbf{x}_n^{\dagger},
\label{eq:xydef}
\end{equation}
while the bath block is always reset to $C_0$. Hence the post-reset SPDM can be parameterized as
\begin{equation}
\rho_n = \begin{pmatrix}
P_n & \mathbf{x}_n\\
\mathbf{y}_n & C_0
\end{pmatrix}.
\label{eq:rhon}
\end{equation}

Let the exact one-cycle propagator be written in block form as
\begin{equation}
U(\tau)=\begin{pmatrix}
u & \mathbf{v}\\
\mathbf{w} & X
\end{pmatrix},
\label{eq:Ublock}
\end{equation}
where $u=u_{00}(\tau)$ is a scalar, $\mathbf{v}=U_{0E}(\tau)$ is a row vector, $\mathbf{w}=U_{E0}(\tau)$ is a column vector, and $X=U_{EE}(\tau)$ is the bath-bath block. The exact pre-reset SPDM is then obtained from Eq.~\eqref{eq:rhonminus} by straightforward block multiplication. The resulting one-cycle map is derived in full detail in Appendix~\ref{app:map}; here we quote the exact component equations:
\begin{align}
P_{n+1} ={}& |u|^2 P_n + u^{*}\mathbf{v}\mathbf{y}_n + u\mathbf{x}_n\mathbf{v}^{\dagger} + \mathbf{v} C_0 \mathbf{v}^{\dagger};
\label{eq:Pmap}\\
\mathbf{x}_{n+1} ={}& \eta\Big[(u P_n + \mathbf{v}\mathbf{y}_n)\mathbf{w}^{\dagger} + (u\mathbf{x}_n + \mathbf{v}C_0)X^{\dagger}\Big];
\label{eq:xmap}\\
\mathbf{y}_{n+1} ={}& \eta\Big[\mathbf{w}(u^{*}P_n + \mathbf{x}_n\mathbf{v}^{\dagger}) + X(\mathbf{y}_nu^{*} + C_0\mathbf{v}^{\dagger})\Big].
\label{eq:ymap}
\end{align}
These equations define an exact affine map
\begin{equation}
\mathbf{W}_{n+1} = A_{\eta}(\tau)\mathbf{W}_n + \mathbf{B}_{\eta}(\tau);
\qquad
\mathbf{W}_n = \big(P_n,\mathbf{x}_n,\mathbf{y}_n\big)^T.
\label{eq:affinemap}
\end{equation}
The fixed point exists and is unique whenever the spectral radius of $A_{\eta}(\tau)$ is smaller than one. In that case,
\begin{equation}
\mathbf{W}_{\fp} = (I-A_{\eta})^{-1}\mathbf{B}_{\eta}.
\label{eq:Wfp}
\end{equation}
Every exact numerical result presented below is obtained from this fixed-point equation. The role of Eqs.~\eqref{eq:Pmap}-\eqref{eq:ymap} is worth spelling out carefully. The occupation $P_n$ does not evolve autonomously once $\eta\neq 0$, because retained coherence feeds back into the system block on the next cycle. Conversely, the coherence block does not simply reflect a passive memory of the previous step: its source term contains the mismatch between the system occupation $P_n$ and the bath occupations encoded in $C_0$; in the bath eigenbasis this appears mode by mode as $P_n-n_k$. The fixed point is therefore self-consistent. One cannot determine the retained coherence without simultaneously determining the system occupation, and one cannot determine the reset heat without knowing the exact pre-reset bath block. This is the basic reason why the coherence-storage problem is nontrivial even in a one-level model.

\subsection{Leading small-$\tau$ asymptotics}

Although the fixed point is computed exactly from Eq.~\eqref{eq:Wfp}, it is useful to derive its leading small-$\tau$ structure analytically. In the bath eigenbasis of $M_{EE}$ and to first order in $\tau$, one finds for the post-reset coherence component $x_{k,n}=\rho_{0k}(t_n^+)$,
\begin{align}
x_{k,n+1} = \eta x_{k,n} &- \ii\eta(\omega_0-\omega_k)x_{k,n}\tau \nonumber\\&+ \ii\eta g_k(P_n-n_k)\tau + \mathcal O(\tau^2).
\label{eq:xkrec}
\end{align}
At fixed point this yields
\begin{equation}
x_k^{\fp}(\tau,\eta)=\frac{\ii\eta g_k\big(P_{\fp}-n_k\big)\tau}{(1-\eta)+\ii\eta(\omega_0-\omega_k)\tau}+\mathcal O(\tau^2),
\label{eq:xkfp}
\end{equation}
and therefore
\begin{equation}
|x_k^{\fp}(\tau,\eta)|^2 = \frac{\eta^2 |g_k|^2\big(P_{\fp}-n_k\big)^2\tau^2}{(1-\eta)^2+\eta^2(\omega_0-\omega_k)^2\tau^2}+\mathcal O(\tau^3).
\label{eq:xkfpmod}
\end{equation}
This expression already shows two qualitative features of the exact theory. First, the overall coherence scale grows with $\eta$. Second, the denominator exhibits a competition between the direct coherence-loss term $(1-\eta)^2$ and a frequency-selective term proportional to $(\omega_0-\omega_k)^2\tau^2$. Thus $\eta$ acts simultaneously as a memory parameter and as a spectral-filter parameter. Appendix~\ref{app:smalltau} gives the derivation step by step.

\section{Structured bath and exact observables}
\label{sec:tb}

\subsection{Semi-infinite tight-binding bath}

To make the structure of the fixed point concrete, we choose a semi-infinite nearest-neighbor chain as the bath,
\begin{equation}
M_{EE} = -J\sum_{j=1}^{\infty}\big(|j\rangle\langle j+1| + |j+1\rangle\langle j|\big),
\label{eq:MEEtb}
\end{equation}
with the system level coupled to the first bath site by amplitude $t_{\rm c}$. The bath eigenmodes are standing waves,
\begin{equation}
\phi_j(k)=\sqrt{\frac{2}{\pi}}\sin(kj),\qquad k\in(0,\pi),
\label{eq:standing}
\end{equation}
with dispersion
\begin{equation}
\omega(k)=-2J\cos k.
\label{eq:dispersion}
\end{equation}
The system-bath coupling in the bath eigenbasis is therefore
\begin{equation}
g(k)=t_{\rm c}\phi_1(k)=t_{\rm c}\sqrt{\frac{2}{\pi}}\sin k.
\label{eq:gk}
\end{equation}
Eliminating $k$ gives the familiar spectral density
\begin{equation}
J(\omega)=\frac{t_{\rm c}^2}{2\pi J^2}\sqrt{4J^2-\omega^2},\qquad |\omega|<2J;
\label{eq:Jomega}
\end{equation}
with $J(\omega)=0$ outside the band. The derivation is given in Appendix~\ref{app:tb}. The band edges at $\omega=\pm 2J$ and the vanishing spectral support outside the band are precisely the ingredients that make this bath a useful structured-environment testbed.

\subsection{Fixed-point retained coherence, reset heat, entropy production, and coherence efficiency}

The exact retained-coherence observable is defined from the fixed-point coherence block as
\begin{equation}
\CSEstar(\tau,\eta)=\sum_k |x_k^{\fp}(\tau,\eta)|^2.
\label{eq:CSEdef}
\end{equation}
When plotted against frequency, the corresponding coherence spectrum is
\begin{equation}
\Scstar(\omega;\tau,\eta)=|x^{\fp}(\omega;\tau,\eta)|^2.
\label{eq:Scomega}
\end{equation}
Equation~\eqref{eq:xkfpmod} then implies the continuum guide formula
\begin{equation}
\Scstar(\omega;\tau,\eta) \approx \frac{\eta^2\tau^2 J(\omega)\big[P_{\fp}-n_F(\omega)\big]^2}{(1-\eta)^2+\eta^2(\omega_0-\omega)^2\tau^2},
\label{eq:Scomeguide}
\end{equation}
where $n_F(\omega)=1/[\e^{\beta(\omega-\mu)}+1]$. The exact numerics are obtained from the full fixed-point map; Eq.~\eqref{eq:Scomeguide} is only a guide to interpretation.

The exact pre-reset bath block at fixed point is the bottom-right block of $U(\tau)\rho_{\fp}U^{\dagger}(\tau)$,
\begin{equation}
C_{\fp}^{-}=\mathbf{w}P_{\fp}\mathbf{w}^{\dagger}+X\mathbf{y}_{\fp}\mathbf{w}^{\dagger}+\mathbf{w}\mathbf{x}_{\fp}X^{\dagger}+XC_0X^{\dagger}.
\label{eq:Cminus}
\end{equation}
The exact reset heat per cycle is defined as the bath energy removed by the super-environment that performs the reset,
\begin{equation}
\Qsupstar(\tau,\eta)=\Tr\big[M_{EE}(C_{\fp}^{-}-C_0)\big],
\label{eq:Qreset}
\end{equation}
and the corresponding exact heat current is
\begin{equation}
\JQstar(\tau,\eta)=\frac{\Qsupstar(\tau,\eta)}{\tau}.
\label{eq:JQdef}
\end{equation}
The exact post-reset system occupation is simply $\Pstar(\tau,\eta)$ from the fixed-point solution.

Finally, to quantify coherence retention relative to thermodynamic price, we define the exact coherence-efficiency ratio
\begin{equation}
R^{*}(\tau,\eta)=\frac{\CSEstar(\tau,\eta)}{\JQstar(\tau,\eta)}.
\label{eq:Rdef}
\end{equation}
For a fixed $\tau$, the three natural operating points are then
\begin{equation}
\eta_{\JQstar}^{\max}(\tau);\qquad \eta_{\CSEstar}^{\max}(\tau);\qquad \eta_{R^{*}}^{\max}(\tau),
\label{eq:etastars}
\end{equation}
obtained by maximizing $\JQstar$, $\CSEstar$, and $R^{*}$, respectively, over $\eta\in[0,1]$. The reset heat current is already sufficient to expose the main tradeoff, but it is useful to emphasize how it relates to the broader thermodynamic framework. If the reference bath state is thermal, the fixed-point entropy production per reset is
\begin{equation}
\Sigmastar(\tau,\eta)=D(\hat\rho_{E,*}^{-}\|\hat\rho_E^{(0)}),
\label{eq:sigmareset}
\end{equation}
and the associated fixed-point entropy-production rate is
\begin{equation}
\sigmastar(\tau,\eta)=\frac{\Sigmastar(\tau,\eta)}{\tau}.
\label{eq:sigmastar_main}
\end{equation}
For Gaussian fermionic states, this quantity is determined by the eigenvalues of the SPDM and can therefore be evaluated directly from $C_{\fp}^{-}$ and $C_0$~\cite{HacklBianchi2021GaussianKahler,Higham2008FunctionsOfMatrices,PtaszynskiEsposito2023GaussianEntropyProduction,PeschelEisler2009FreeLatticeRDM}. We do not plot $\sigmastar$ in the main figures, because the present manuscript is already fully resolved by the heat-current tradeoff.

\section{Exact results: coherence spectra, tradeoff curves, and operating diagrams}
\label{sec:results}

\begin{figure}[!htb]
\centering
\includegraphics[width=0.9\columnwidth]{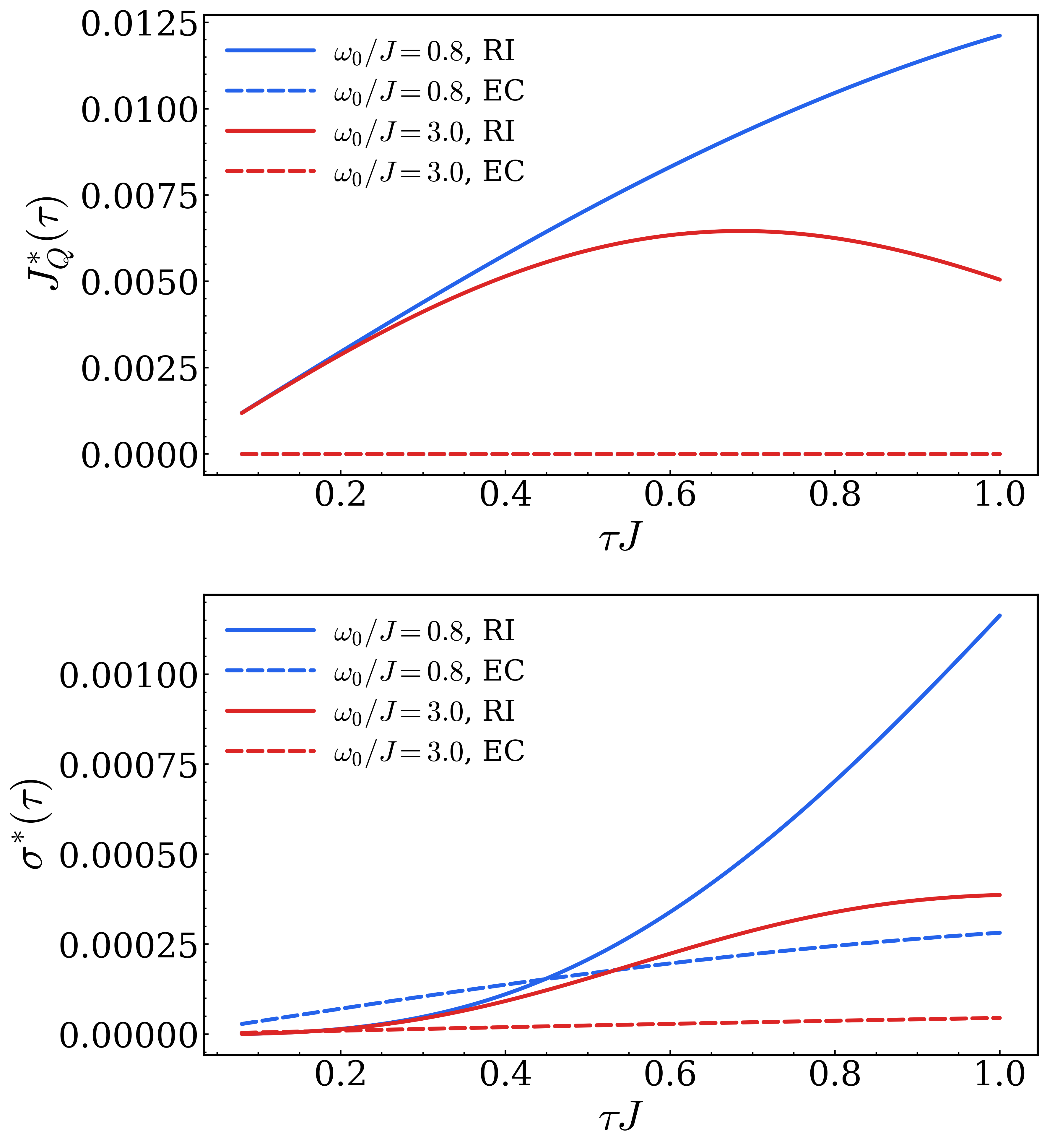}
\caption{Exact baseline thermodynamic comparison for the endpoint reset channels. Upper panel: fixed-point reset heat current $J_Q^{*}(\tau)=Q_{\mathrm{sup}}^{*}(\tau)/\tau$. Lower panel: fixed-point entropy-production rate $\sigma^{*}(\tau)=\Sigma_{\mathrm{reset}}^{*}(\tau)/\tau$. Solid curves correspond to the repeated-interaction (RI) endpoint, $\eta=0$, while dashed curves correspond to the evolving-correlation (EC) endpoint, $\eta=1$. Blue curves show the in-band level $\omega_0/J=0.8$, and red curves show the out-of-band level $\omega_0/J=3.0$. Here $\mu=0$, and we use units with $k_B=1$, so $\Sigma_{\mathrm{reset}}^{*}$ is dimensionless and $\sigma^{*}$ is measured in units of $k_{\rm B}$ per unit time.}
\label{fig:thermo-baseline}
\end{figure}
Figure~\ref{fig:thermo-baseline} summarizes the exact fixed-point thermodynamics of the two endpoint reset channels.
The upper panel shows the fixed-point reset heat current $J_Q^{*}(\tau)$, while the lower panel shows the fixed-point entropy-production rate $\sigma^{*}(\tau)$. Solid curves correspond to the repeated-interaction endpoint $\eta=0$, while dashed curves correspond to the evolving-correlation endpoint $\eta=1$. For each channel we compare an inside-band level, $\omega_0/J=0.8$, and an outside-band level, $\omega_0/J=3.0$. Several features are immediately clear. First, the in-band configuration is thermodynamically much more active than the out-of-band one. When $\omega_0$ lies inside the bath continuum, the system hybridizes efficiently with the bath modes, and both the steady reset heat current and the entropy production are enhanced. When $\omega_0$ lies outside the band, the exchange with the continuum is effectively off-resonant, so the thermodynamic response is substantially reduced.

Second, the two endpoint channels differ qualitatively. At the RI endpoint, the bath block after the unitary stroke differs appreciably from its reset reference state, producing a visible heat current and a correspondingly larger entropy-production rate. At the EC endpoint, the reset acts only on the bath while preserving the propagated system--environment coherences. As a result, the bath-energy cost vanishes at the fixed point, 
\begin{equation}
   \Qsupstar(\tau,\eta=1)=0; \quad \JQstar(\tau,\eta=1)=0, 
\end{equation}
even though the entropy production need not vanish because the pre-reset bath state can still differ from the reference state at the level of its full Gaussian structure. This is why the dashed EC heat-current curves in Fig.~\ref{fig:thermo-baseline} sit at zero, while the corresponding entropy-production curves remain finite. Figure~\ref{fig:thermo-baseline} therefore provides the natural thermodynamic baseline against which the coherence-selective protocol should be compared. The central question of the present work is how the partial retention of system-environment coherence modifies this exact fixed-point thermodynamic benchmark.

\begin{figure*}[t]
\centering
\includegraphics[width=0.48\textwidth]{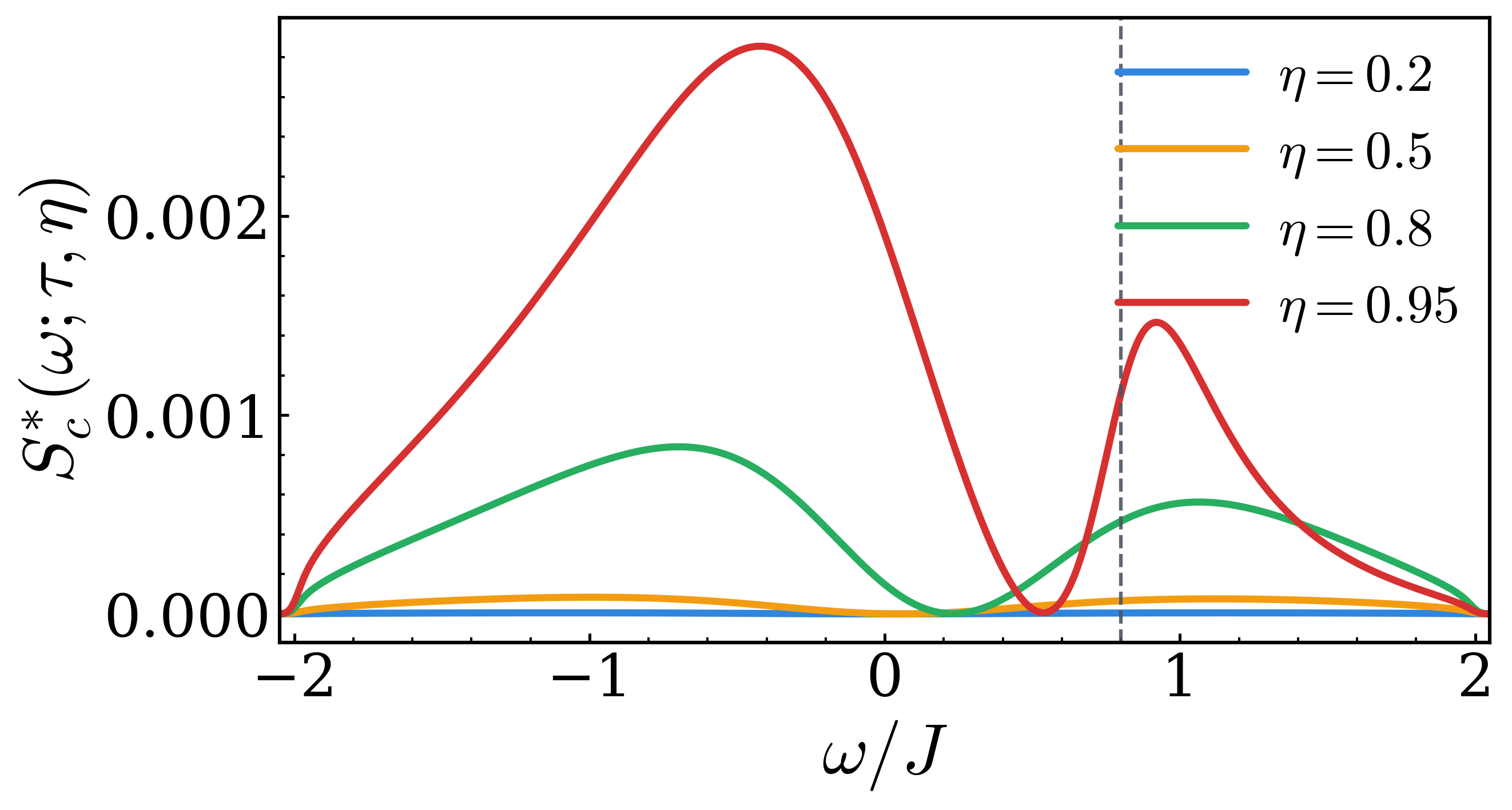}
\includegraphics[width=0.48\textwidth]{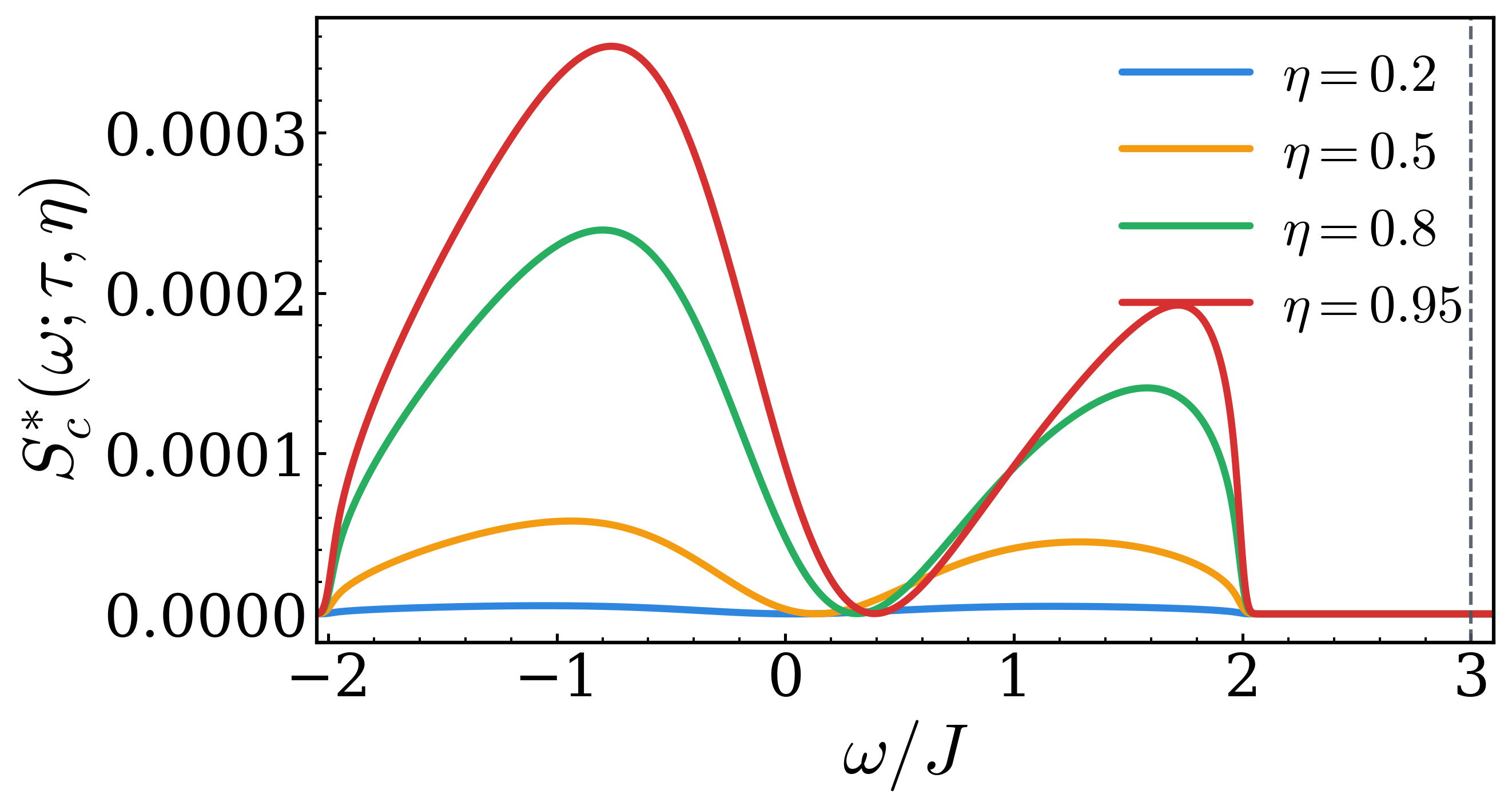}
\caption{Exact fixed-point retained-coherence spectra $S_c^{*}(\omega;\tau,\eta)$ for the coherence-selective reset channel at fixed stroboscopic interval $\tau J=0.2$. Left: inside-band level, $\omega_0/J=0.8$. Right: outside-band level, $\omega_0/J=3.0$. In each panel the curves correspond to $\eta=0.2,\,0.5,\,0.8,$ and $0.95$, and the vertical dashed line marks the bare system level $\omega_0$. For the inside-band case, the retained coherence is much larger because the level hybridizes with resonant bath modes inside the tight-binding continuum. For the outside-band case, where no bath modes exist at $\omega_0$, the coherence cannot accumulate on shell and is instead redistributed over the allowed bath band, with weight pushed back toward the band edge. Thus $\eta$ controls both the magnitude and the spectral localization of the fixed-point system--environment coherence, while the structured bath determines where that coherence is supported in frequency space.}
\label{fig:spectra}
\end{figure*}

All results are obtained from the exact fixed-point solution of Eq.~\eqref{eq:Wfp} for a large truncated chain. Unless otherwise stated, we use $J=1$, $t_{\rm c}=0.2$, $\beta=3$, and $\mu=0$. Convergence with bath size is discussed in Appendix~\ref{app:numerics}. The line-cut figures at fixed $\tau$ use $\tau J=0.2$, while the heat maps sweep both $\tau$ and $\eta$. Figure~\ref{fig:spectra} shows the exact retained-coherence spectrum for an inside-band level, $\omega_0/J=0.8$, and an outside-band level, $\omega_0/J=3.0$. Several features deserve emphasis. In all plots we use units with $k_{\rm B}=1$.

\begin{figure*}[t]
\centering
\includegraphics[width=0.48\textwidth]{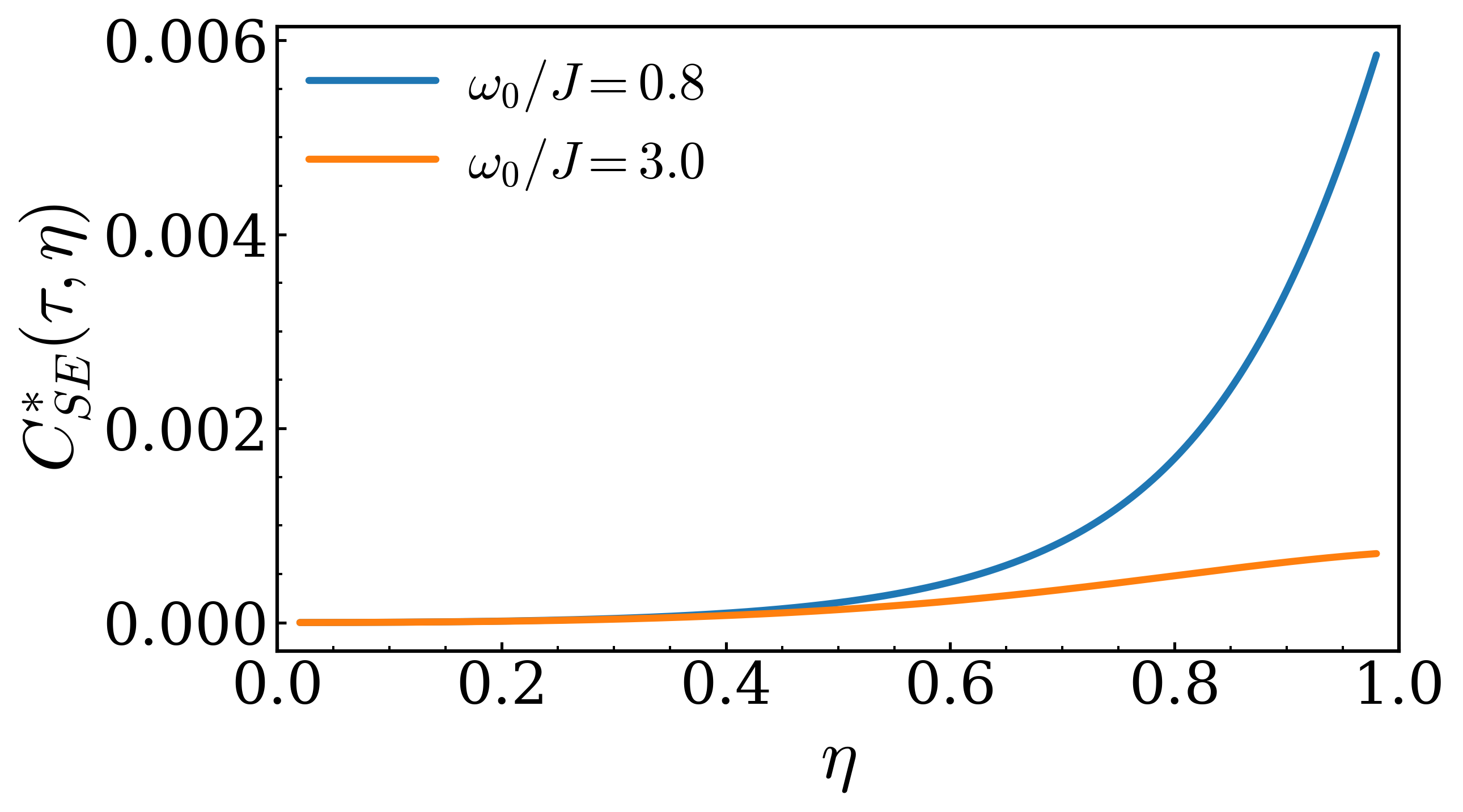}
\includegraphics[width=0.48\textwidth]{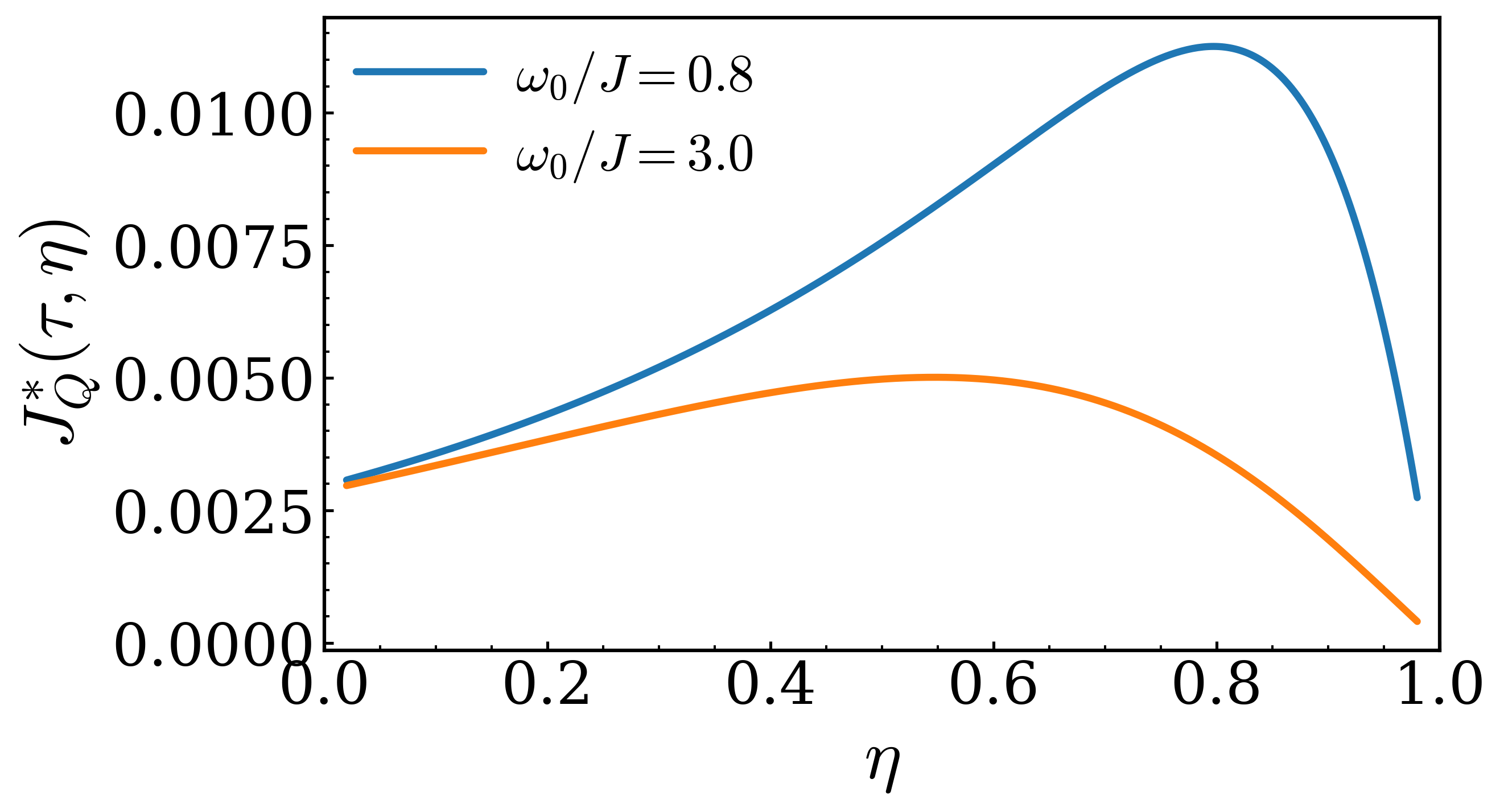}
\caption{Exact fixed-$\tau$ cuts at $\tau J=0.2$ for the coherence-selective reset channel as a function of the retention parameter $\eta$. Left panel: total retained system--environment coherence, \(
C_{SE}^{*}(\tau,\eta) = \int d\omega\, S_c^{*}(\omega;\tau,\eta).\) Right panel: fixed-point reset heat current, \(J_Q^{*}(\tau,\eta)=Q_{\mathrm{sup}}^{*}(\tau,\eta)/\tau.\) Blue curves correspond to the inside-band level $\omega_0/J=0.8$, while orange curves correspond to the outside-band level $\omega_0/J=3.0$. The retained coherence increases strongly with $\eta$, with a much larger scale for the inside-band case because resonant hybridization with the bath continuum allows coherence to accumulate efficiently. By contrast, the reset heat current is generically nonmonotonic in $\eta$: partial retention of system--environment coherence can initially enhance the thermodynamic response, but as $\eta\to1$ the reset action becomes weaker and the per-cycle heat current is reduced. 
This figure therefore shows that coherence retention and thermodynamic throughput are related but not equivalent: increasing retained memory does not imply a monotonic increase of the reset heat current.}
\label{fig:linecuts}
\end{figure*}
First, for both level positions the spectrum grows strongly with $\eta$, in agreement with the analytic expectation from Eq.~\eqref{eq:xkfpmod}. Second, the in-band spectrum is much larger than the out-of-band spectrum. This reflects the straightforward physical fact that an in-band level hybridizes with resonant bath modes, while an out-of-band level must build its coherence through off-resonant dressing of the band. Third, the spectral support is highly nontrivial. In the inside-band case, the largest retained coherence does not simply sit at the bare level position; instead it is redistributed across the available bath support. In the outside-band case, where no bath states exist at $\omega_0$, the coherence is entirely pushed back onto the band. Thus the fixed-point memory is structured not by the bare level alone, but by the full bath support and by the exact one-cycle propagator.

The simplest way to see the central result is to fix the stroboscopic interval and vary $\eta$. Relative to the endpoint thermodynamic baselines of Fig.~\ref{fig:thermo-baseline}, Figure~\ref{fig:linecuts} collects the exact total retained coherence $\CSEstar(\tau,\eta)$ and the exact reset heat current $\JQstar(\tau,\eta)$. The retained coherence is monotonic in $\eta$ for both level positions shown. By contrast, the heat current is not: in both cases it rises at small $\eta$, reaches a clear interior maximum, and then falls again as $\eta\to1$. The out-of-band case exhibits the same qualitative structure, although on a smaller coherence scale. Taken together, these line cuts already establish the central fixed-point statement of this work: the coherence-preserving reset channel is not the heat-maximizing reset channel.

\begin{figure*}[t]
\centering
\includegraphics[width=0.48\textwidth]{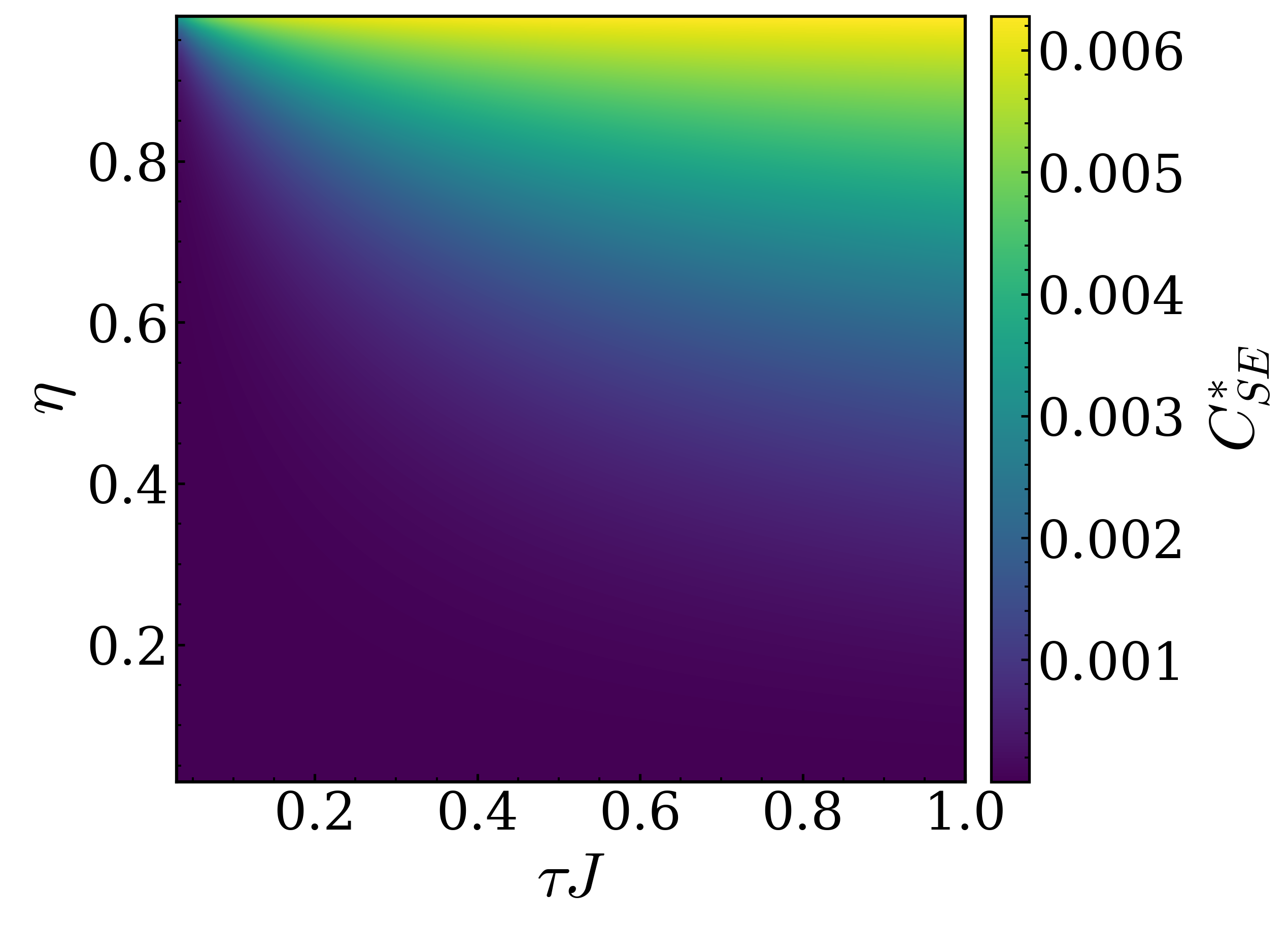}
\includegraphics[width=0.48\textwidth]{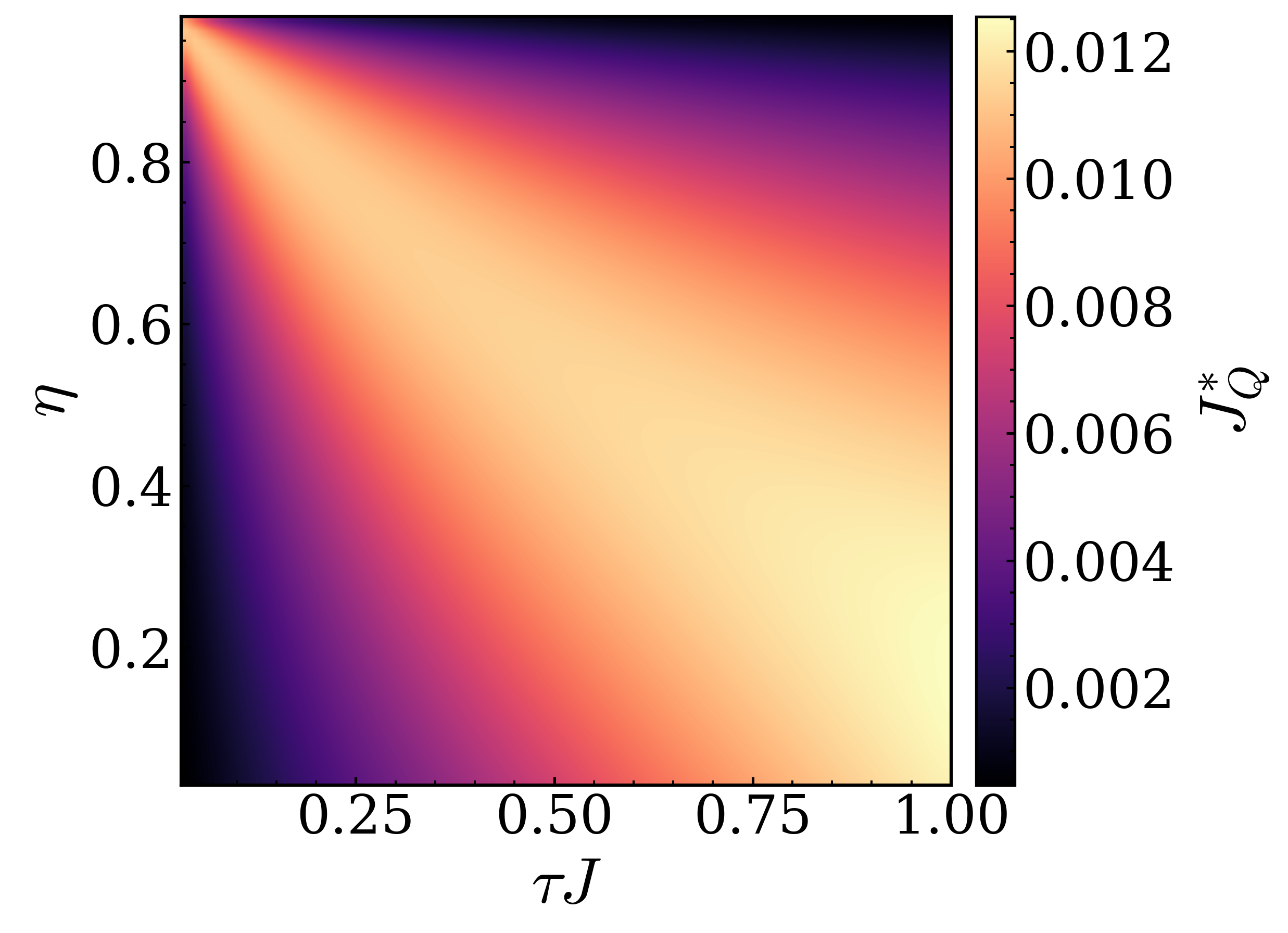}\\
\includegraphics[width=0.48\textwidth]{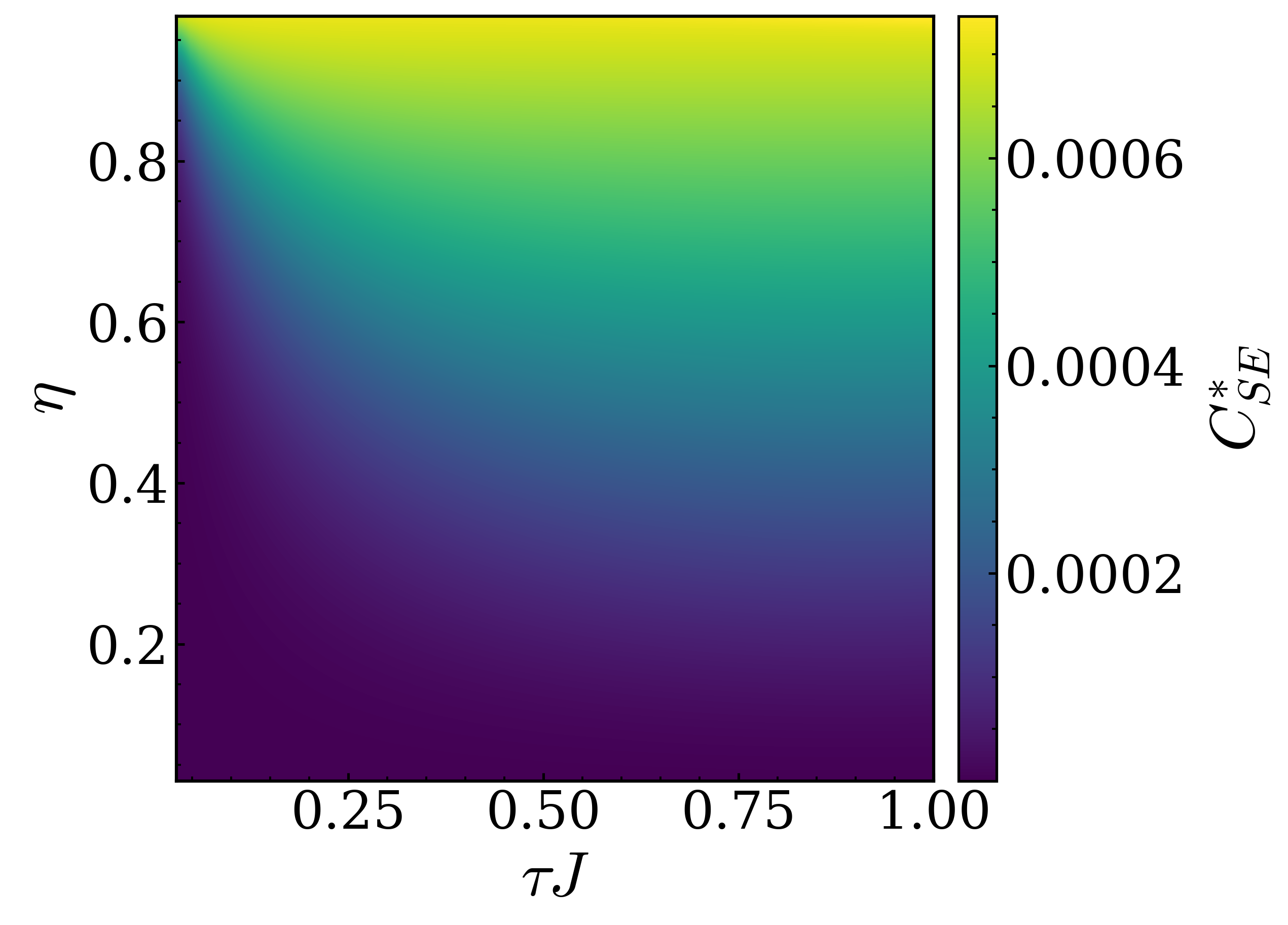}
\includegraphics[width=0.48\textwidth]{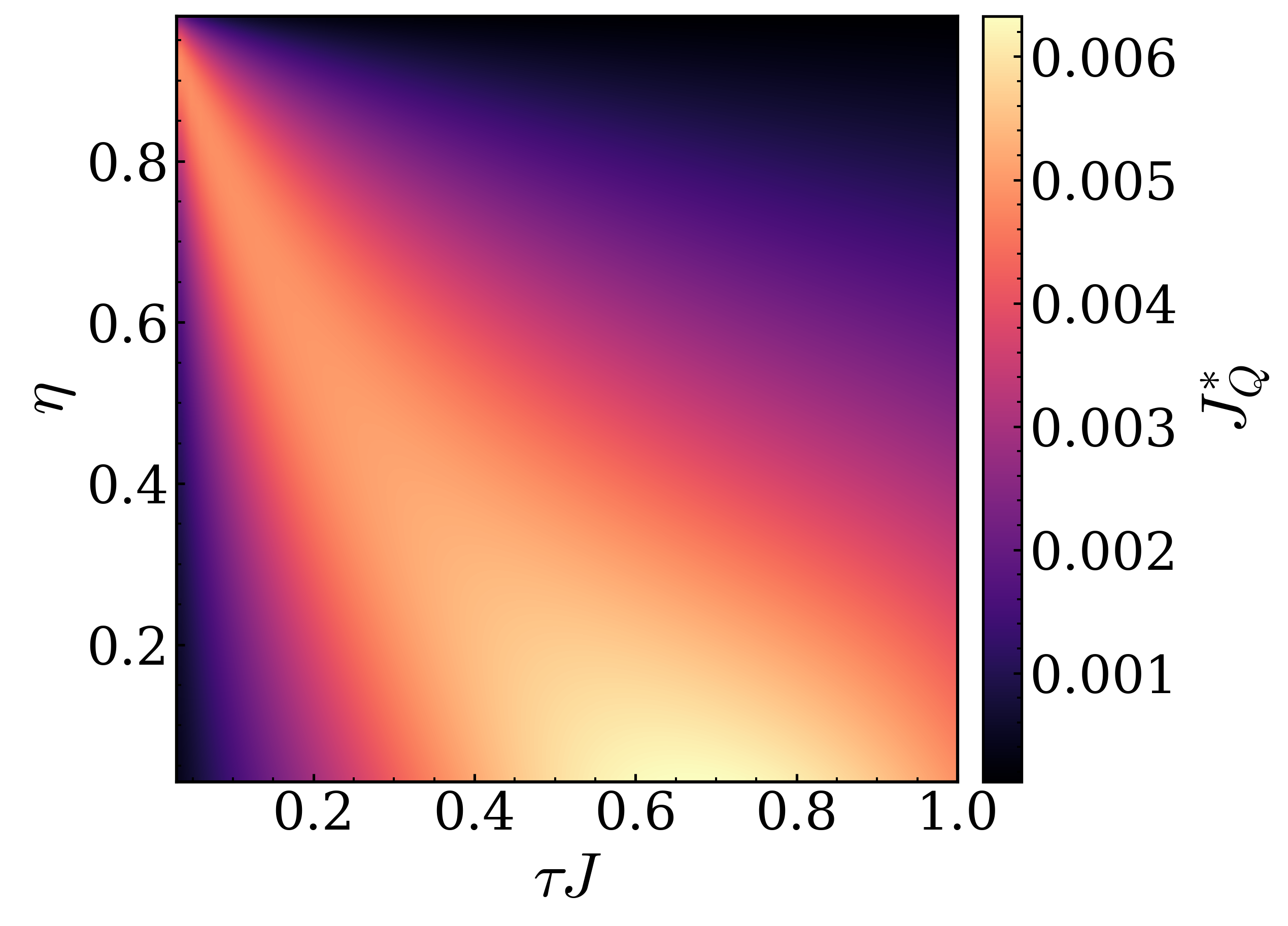}
\caption{Exact fixed-point landscapes for the coherence-selective reset channel in the $(\tau,\eta)$ plane. Upper row: inside-band level $\omega_0/J=0.8$. Lower row: outside-band level $\omega_0/J=3.0$. Left column: total retained system--environment coherence $C_{SE}^{*}(\tau,\eta)$. Right column: fixed-point reset heat current $J_Q^{*}(\tau,\eta)=Q_{\mathrm{sup}}^{*}(\tau,\eta)/\tau$. In both band placements, the retained coherence grows strongly as $\eta\to1$, since a larger fraction of the propagated system--environment coherence is preserved from cycle to cycle; it also increases with $\tau$ as the unitary stroke has more time to build up correlations. By contrast, the reset heat current is nonmonotonic over the $(\tau,\eta)$ plane and develops a broad high-response ridge rather than simply maximizing at the coherence-preserving endpoint. The inside-band case exhibits a much larger overall scale in both observables because the system level hybridizes efficiently with bath modes inside the tight-binding continuum.
For the outside-band case, the same qualitative structure survives on a compressed scale, reflecting the absence of resonant bath support at the bare level energy. The figure shows that maximal retained coherence and maximal thermodynamic throughput do not coincide: coherence accumulation is largest near $\eta=1$, whereas heat extraction is optimized in an extended intermediate region where coherence retention and reset action are balanced.}
\label{fig:heatmaps-in}
\end{figure*}

Figure~\ref{fig:heatmaps-in} displays the exact fixed-point landscapes of the coherence-selective reset channel over the full $(\tau,\eta)$ plane for both an inside-band and an outside-band system level. The upper row corresponds to $\omega_0/J=0.8$, where the level lies inside the tight-binding continuum, while the lower row corresponds to $\omega_0/J=3.0$, where the level lies outside the bath band. In each row, the left panel shows the total retained system-environment coherence $C_{SE}^{*}(\tau,\eta)$ and the right panel shows the fixed-point reset heat current $J_Q^{*}(\tau,\eta)$.

Several features are immediately visible. First, in both band placements the retained coherence increases strongly with the retention parameter $\eta$ and also grows with the reset interval $\tau$. This is physically natural: larger $\eta$ means that a greater fraction of the propagated system-environment coherence survives each reset step, while larger $\tau$ allows the unitary segment to generate stronger correlations before the next reset is applied. Second, the inside-band case supports a much larger retained-coherence scale than the outside-band case, because an in-band level can hybridize efficiently with resonant bath modes, whereas an out-of-band level cannot accumulate coherence on shell with the same effectiveness.

The heat-current landscapes reveal a qualitatively different structure. In both the inside-band and outside-band cases, $J_Q^{*}(\tau,\eta)$ is nonmonotonic in $\eta$ and forms a broad high-response ridge across the $(\tau,\eta)$ plane. Thus, the thermodynamically optimal regime is not the coherence-preserving endpoint itself. Instead, the largest reset heat current is achieved in an extended intermediate region where coherence retention is substantial, but the reset action remains strong enough to sustain an appreciable per-cycle energy exchange with the super-environment. This difference between the coherence landscape and the heat-current landscape makes the coherence-cost tradeoff visually transparent: the protocol that stores the most coherence is not the protocol that dissipates the most heat.

Comparing the two rows further shows that band placement controls the overall thermodynamic scale without changing the basic geometry of the tradeoff. The outside-band case retains the same qualitative pattern as the inside-band case, but both $C_{SE}^{*}$ and $J_Q^{*}$ are reduced because the system level is spectrally detached from the bath continuum. The four panels together therefore provide the most complete picture of the fixed-point coherence-cost structure of the reset channel: the retention parameter $\eta$ governs memory preservation, the reset interval $\tau$ controls the amount of coherence generated during each cycle, and the bath spectral support determines the absolute strength of both coherence buildup and thermodynamic cost.

\begin{figure}[t]
\centering
\includegraphics[width=0.48\textwidth]{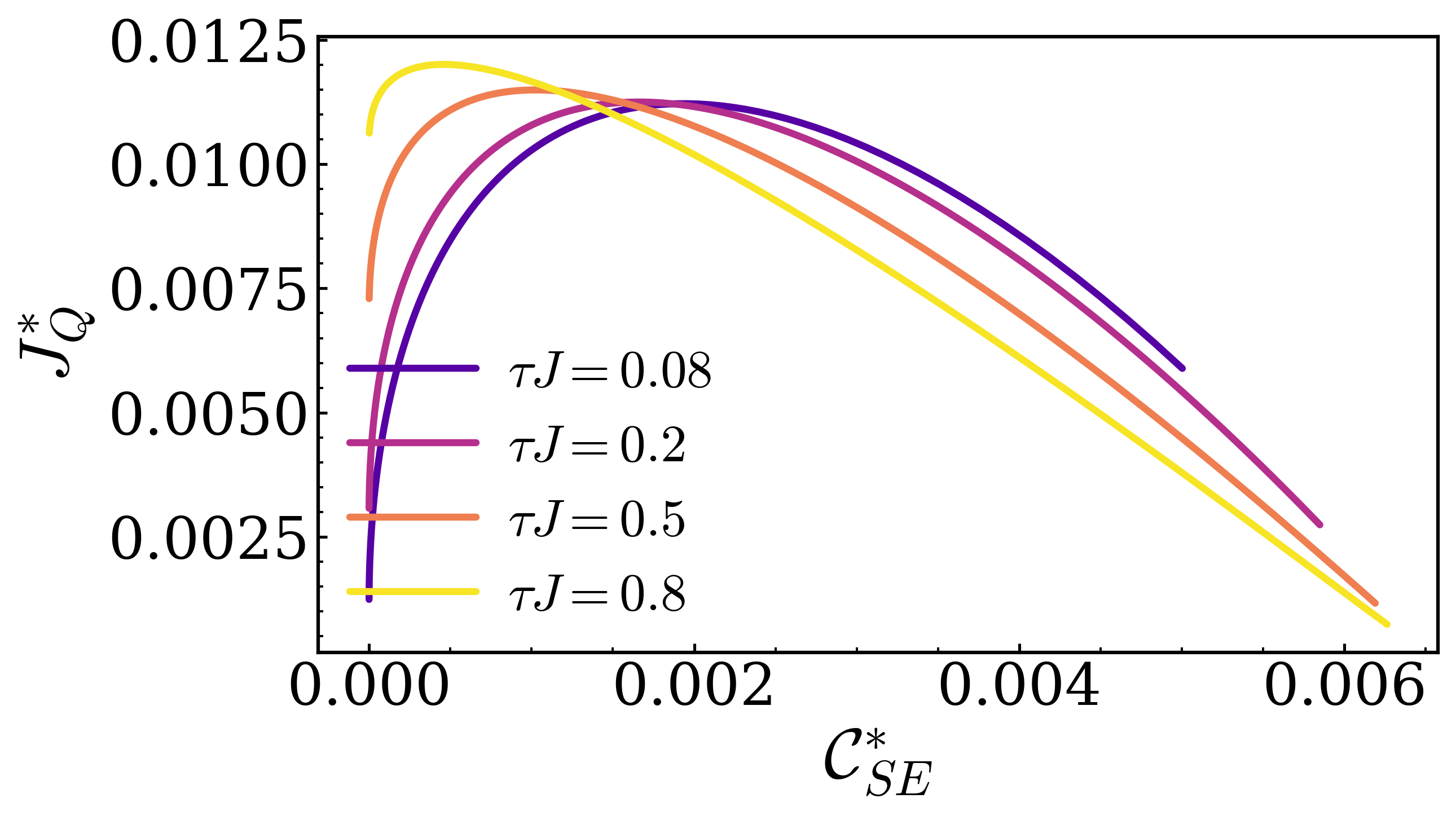}
\includegraphics[width=0.48\textwidth]{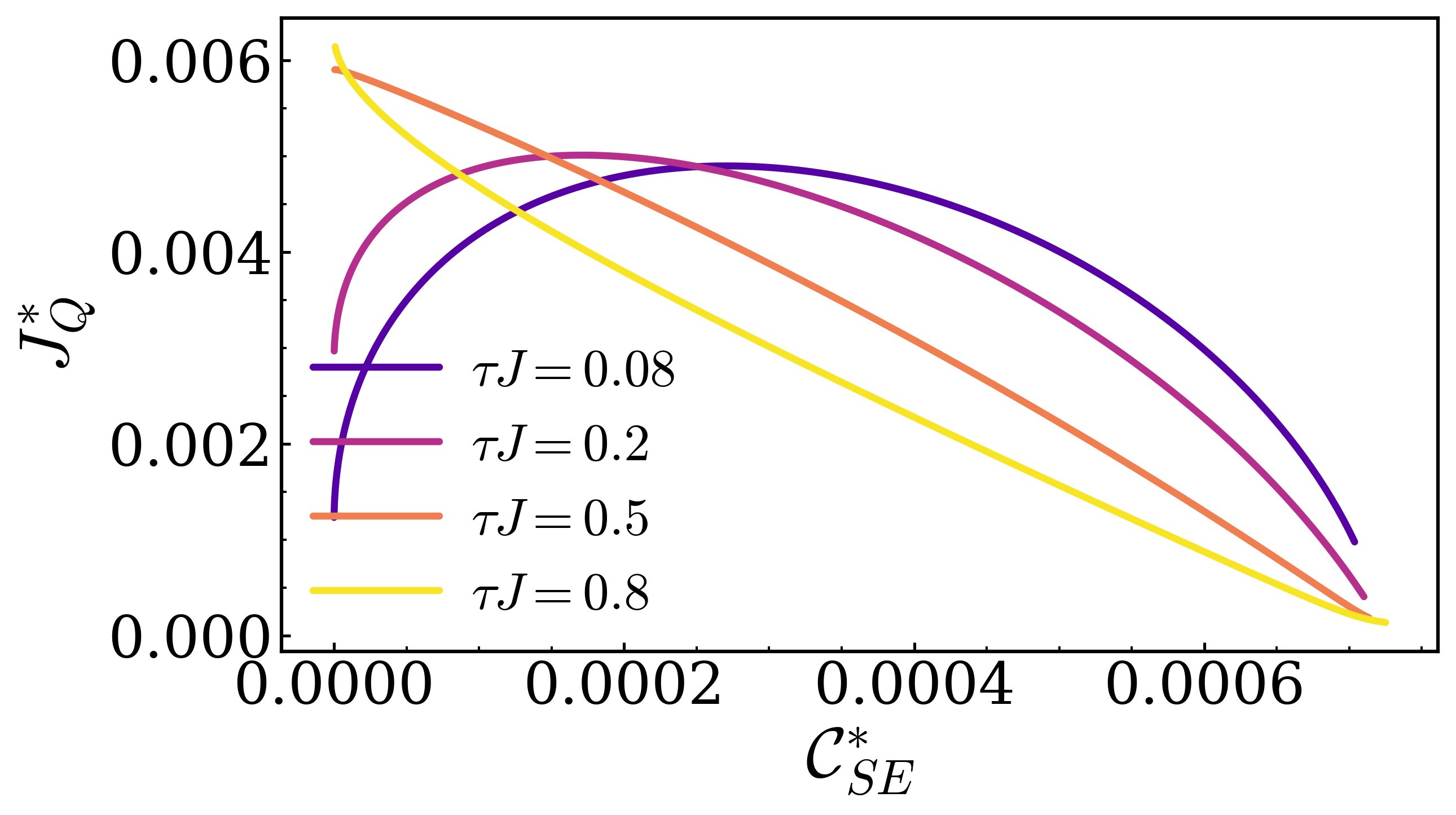}
\caption{Exact Pareto-style operating diagrams in the $(C_{SE}^{*},J_Q^{*})$ plane for the coherence-selective reset channel, obtained by sweeping $\eta$ at fixed $\tau$. Upper panel: inside-band level $\omega_0/J=0.8$. Lower panel: outside-band level $\omega_0/J=3.0$. Each colored curve corresponds to a fixed reset interval $\tau J=0.08,\,0.2,\,0.5,$ and $0.8$. In both band placements, the curves form an arch-like tradeoff geometry that separates a cost-building branch from a cost-relieving branch: as $\eta$ is increased from the RI endpoint, the reset heat current initially rises together with the retained coherence, but beyond an interior optimum, the heat current decreases while the retained coherence continues to grow toward the coherence-preserving endpoint. Thus, larger retained coherence does not require larger reset heat all the way to $\eta=1$. The inside-band case spans a much larger coherence and heat-current scale because the system level hybridizes resonantly with bath modes in the tight-binding continuum. The outside-band case exhibits the same qualitative geometry on a compressed scale, reflecting the absence of resonant bath support at the bare level energy.}
\label{fig:pareto}
\end{figure}

These operating diagrams in Figure~\ref{fig:pareto} make the coherence-cost tradeoff particularly transparent. For each fixed $\tau$, the trajectory is not a monotone graph but an arch-like curve. Starting from the RI side, increasing $\eta$ initially increases both the retained coherence and the reset heat current, because partial preservation of system-environment coherence enhances the memory stored across cycles while still allowing the reset to act strongly on the bath block. However, once $\eta$ becomes too large, the reset action weakens: the retained coherence continues to increase, but the per-cycle heat current decreases. As a result, the heat-optimal operating point lies at an interior value of $\eta$, whereas the coherence-optimal operating point is pushed toward the coherence-preserving endpoint. The comparison between the two panels shows that band placement changes the overall thermodynamic scale without changing the basic tradeoff geometry. For the inside-band level, resonant hybridization with the bath continuum allows both larger fixed-point coherence and larger reset heat current. For the outside-band level, the same qualitative structure survives but on a much smaller scale, since the level lies outside the bath band and cannot exchange coherence and energy with resonant bath modes in the same way. The operating diagrams therefore summarize the central message of the manuscript in a compact way: coherence retention, thermodynamic cost, and spectral placement are tightly linked, but the protocol that stores the most coherence is not the protocol that dissipates the most heat.

\begin{figure*}[t]
\centering
\includegraphics[width=0.48\linewidth]{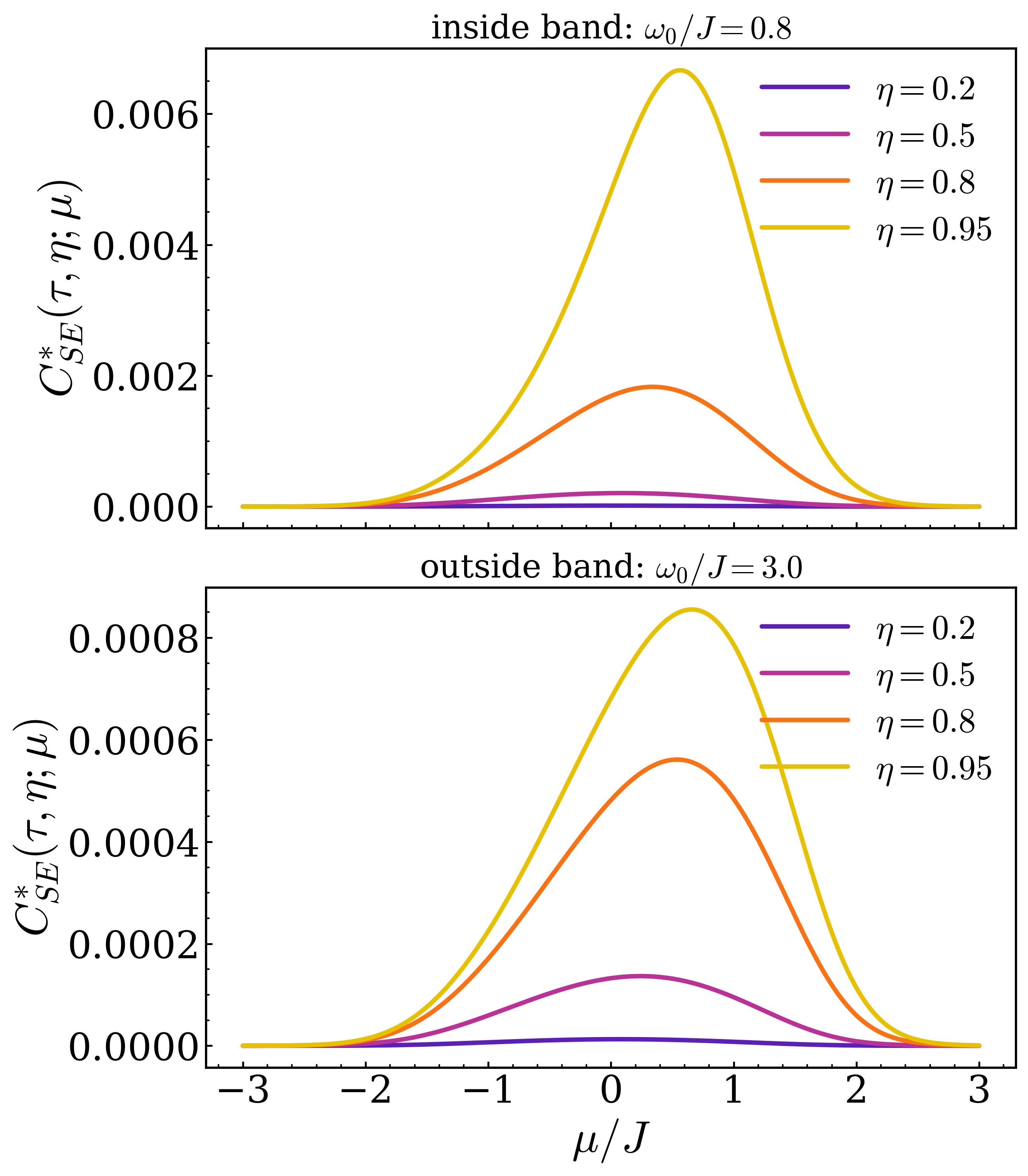}\hfill
\includegraphics[width=0.48\linewidth]{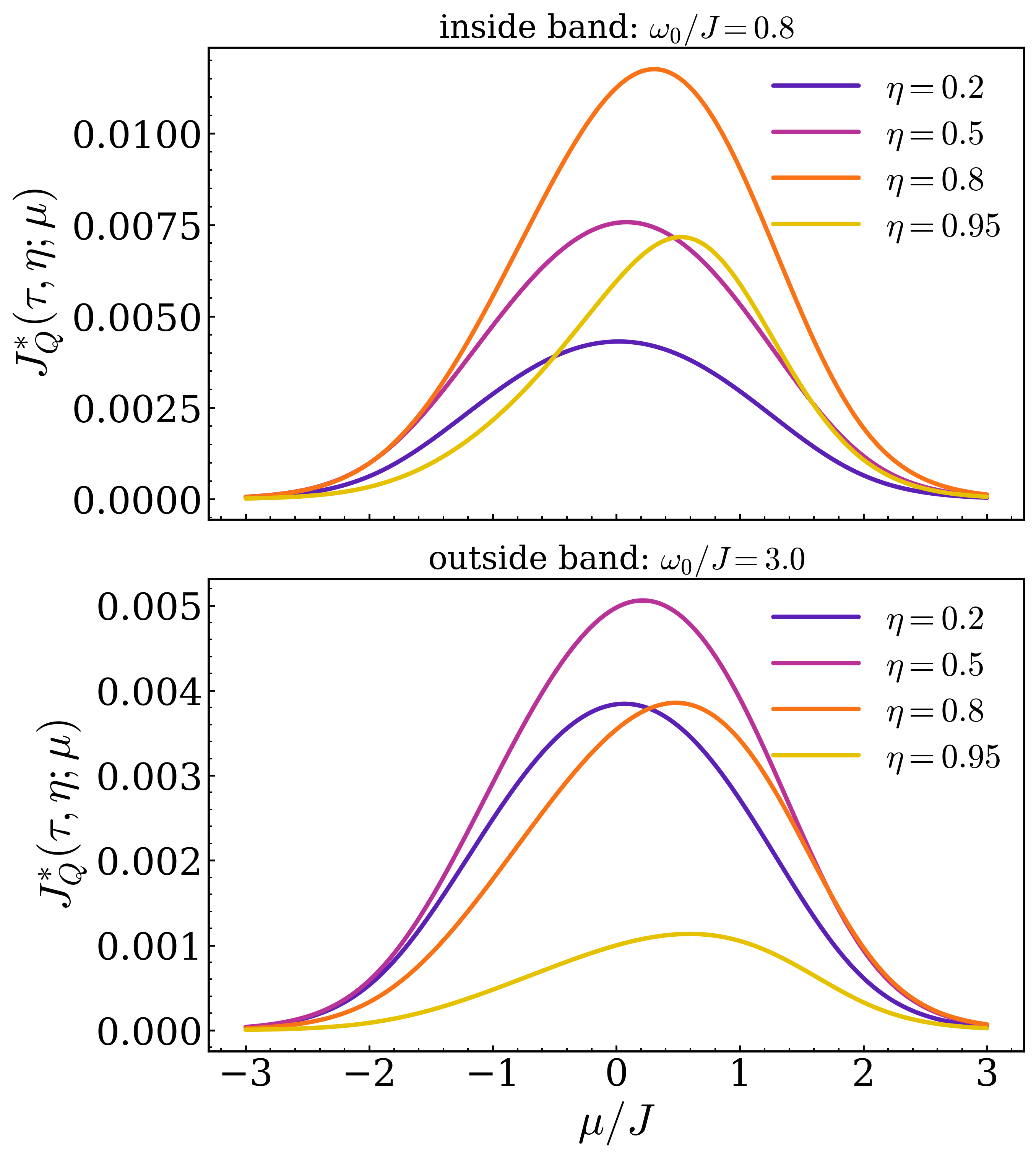}
\caption{Exact $\mu$-sweep fixed-point observables at $\tau J=0.2$ for the coherence-selective reset channel.
Left: retained coherence $C_{SE}^{*}(\tau,\eta,\mu)$. Right: reset heat current $J_Q^{*}(\tau,\eta,\mu)$.
In each panel, the upper subplot corresponds to the inside-band level $\omega_0/J=0.8$ and the lower subplot to the outside-band level $\omega_0/J=3.0$. The curves correspond to $\eta=0.2,\,0.5,\,0.8,$ and $0.95$. Both observables are strongly nonmonotonic in $\mu$, and their maxima drift toward larger positive chemical potential as $\eta$ increases.
The inside-band case shows a much larger scale than the outside-band case, but the same qualitative filling-biased structure survives outside the band on a compressed scale.}
\label{fig:mu_sweep_CSE_JQ}
\end{figure*}

\section{Filling-biased reset targets: nonzero chemical potential}
\label{sec:mu_extension}

The previous sections focused on the particle-hole-symmetric reference choice $\mu=0$. It is natural to ask whether a nonzero chemical potential produces qualitatively new fixed-point behavior. In the present reset protocol, the chemical potential enters through the reference bath state,
\begin{equation}
C_0(\mu)=\mathrm{diag}\!\left[n_F(\omega_k;\mu)\right];
\quad
n_F(\omega;\mu)=\frac{1}{e^{\beta(\omega-\mu)}+1},
\label{eq:C0_mu_ext}
\end{equation}
and therefore changes both the fixed-point system occupation $P_{\mathrm{fp}}$ and the mismatch term $P_{\mathrm{fp}}-n_F(\omega;\mu)$ that drives the retained coherence. Indeed, the small-$\tau$ guide formula of Eq.~(\ref{eq:Scomeguide}) becomes
\begin{equation}
S_c^{*}(\omega;\tau,\eta,\mu)\approx
\frac{\eta^2\tau^2 J(\omega)\left[P_{\mathrm{fp}}(\tau,\eta,\mu)-n_F(\omega;\mu)\right]^2}
{(1-\eta)^2+\eta^2(\omega_0-\omega)^2\tau^2},
\label{eq:Sc_mu_ext}
\end{equation}
so a nonzero $\mu$ changes not only the overall scale of the fixed-point coherence but also its spectral mismatch profile.

A priori, one might therefore expect a new grand-canonical cost structure. Define the fixed-point particle transfer into the super-environment during one reset by
\begin{align}
\Delta N_E^{*}(\tau,\eta,\mu)&=\Tr\!\left[C_{\mathrm{fp}}^{-}(\tau,\eta,\mu)-C_0(\mu)\right];
\\
J_N^{*}(\tau,\eta,\mu)&=\frac{\Delta N_E^{*}(\tau,\eta,\mu)}{\tau},
\label{eq:JN_def_ext}
\end{align}
and the corresponding grand-canonical reset-cost current by
\begin{equation}
J_{\mathrm{gc}}^{*}(\tau,\eta,\mu)
=
\frac{Q_{\mathrm{sup}}^{*}(\tau,\eta,\mu)-\mu\,\Delta N_E^{*}(\tau,\eta,\mu)}{\tau}.
\label{eq:Jgc_def_ext}
\end{equation}
However, for the present protocol this apparent generalization simplifies exactly at the stroboscopic fixed point.
The reset leaves the system block unchanged, so the pre-reset system occupation at the end of a cycle is identical to the next post-reset occupation. Since the unitary segment conserves the total particle number,
\begin{equation}
P_{\mathrm{fp}}+\Tr_E C_0
=
P_{\mathrm{fp}}^{-}+\Tr_E C_{\mathrm{fp}}^{-},
\label{eq:Ncons_ext}
\end{equation}
and at the fixed point one has $P_{\mathrm{fp}}^{-}=P_{\mathrm{fp}}$.
Therefore
\begin{equation}
\Delta N_E^{*}(\tau,\eta,\mu)=0,
\qquad
J_N^{*}(\tau,\eta,\mu)=0,
\label{eq:JN_zero_ext}
\end{equation}
exactly, and hence
\begin{equation}
J_{\mathrm{gc}}^{*}(\tau,\eta,\mu)=J_Q^{*}(\tau,\eta,\mu).
\label{eq:Jgc_equals_JQ_ext}
\end{equation}
Thus a nonzero chemical potential does not create an independent fixed-point particle-current cost in this protocol.
Its role is instead to bias the reset target $C_0(\mu)$ and thereby reshape the fixed-point coherence, the fixed-point system occupation, and the reset heat current.

Figure~\ref{fig:mu_sweep_CSE_JQ} shows the resulting $\mu$-sweep line cuts at fixed $\tau J=0.2$ for four representative values of $\eta$. The upper row corresponds to the inside-band level $\omega_0/J=0.8$, while the lower row corresponds to the outside-band level $\omega_0/J=3.0$. Several clear conclusions emerge. First, both $C_{SE}^{*}$ and $J_Q^{*}$ are strongly nonmonotonic as functions of $\mu$. Second, the maxima are displaced to positive $\mu$, and this displacement becomes larger as $\eta$ increases. Third, the inside-band case remains parametrically larger than the outside-band case, but the outside-band level exhibits the same qualitative filling-biased structure on a compressed scale. In this sense, nonzero $\mu$ acts as a genuine control knob for the coherence-cost geometry,
even though the fixed-point particle current itself vanishes.

The nonzero-$\mu$ behavior originates from the way the chemical potential reshapes the reset target occupations entering the exact fixed point. Sweeping $\mu$ across the bath band shifts the Fermi profile $n_F(\omega;\mu)$ and therefore modifies the self-consistent mismatch $P_{\mathrm{fp}}-n_F(\omega;\mu)$ that drives both the retained coherence and the reset heat current. As a result, the maxima of $C_{SE}^{*}$ and $J_Q^{*}$ move systematically with $\mu$ and with the retention parameter $\eta$. Thus the $\mu\neq0$ extension is best understood as a filling-biased deformation of the original coherence-heat tradeoff, not as a new three-way coherence--heat--particle tradeoff at the
fixed point.

This observation also clarifies the appropriate figure of merit. Because Eq.~\eqref{eq:Jgc_equals_JQ_ext} holds exactly, the grand-canonical coherence-per-cost ratio
\begin{equation}
R_{\mu}^{*}(\tau,\eta,\mu)
=
\frac{C_{SE}^{*}(\tau,\eta,\mu)}{J_{\mathrm{gc}}^{*}(\tau,\eta,\mu)}
\label{eq:Rmu_def_ext}
\end{equation}
coincides identically with $C_{SE}^{*}/J_Q^{*}$ at the fixed point. Thus the main new physics of $\mu\neq 0$ is not the appearance of an additional transport channel, but the way in which filling bias shifts and reshapes the exact coherence and heat landscapes. This makes nonzero $\mu$ a natural extension of the present work, but also shows that it is best understood as a filling-controlled continuation of the same exact reset geometry rather than as a
fundamentally different thermodynamic regime.

\section{Summary and conclusions}
\label{sec:discussion}

In this work, we have considered the thermodynamic consequences of a class of reset protocols that affect the system-environment coherences within the SPDM picture of quadratic Hamiltonians. The exact fixed-point solution shows that the parameter $\eta$ interpolating between complete erasure and retainment of SPDM coherence blocks is a genuine control knob for the cycle-stationary memory-cost geometry. It is thus a genuine control parameter for memory engineering in a structured bath. The reason is simple but fundamental. The reset channel does two logically distinct things at once. It reinitializes the bath occupations, and it erases only part of the system-environment coherence. Those two actions need not be optimized by the same operating point.

This leads to four concrete physical conclusions. First, $\eta$ controls exact memory retention. The total retained coherence $\CSE(\tau,\eta)$ is monotone in $\eta$ throughout the numerical data set considered here. The coherence-preserving endpoint is therefore the natural memory-maximizing limit. Second, the thermodynamic price of resetting is not monotone in memory retention. The exact reset heat current $\JQstar(\tau,\eta)$ is largest at an intermediate $\eta$, not at $\eta=1$. Intuitively, if $\eta$ is too small, the channel erases coherence so aggressively that very little memory survives from cycle to cycle. If $\eta$ is too large, the channel erases too little at each reset to generate a maximal heat dump into the super-environment. Maximal thermodynamic burden therefore occurs at an intermediate balance between coherence creation during the unitary step and coherence erasure during the reset. Third, coherence efficiency is pushed toward the near-$\EC$ regime. The ratio $R^{*}=\CSEstar/\JQstar$ is optimized near the coherence-preserving endpoint. In other words, the same sector that stores the most memory is also the sector that stores memory most efficiently per unit reset heat, consistent with the finding that pure decoherence necessarily dissipates heat into the environment \cite{Popovic2023}. Fourth, the bath structure matters throughout. The in-band case supports much larger memory because resonant bath modes are available at the level energy. The outside-band case suppresses the overall coherence scale but preserves the same tradeoff geometry. The effect is therefore not a trivial artifact of the interpolation parameter. It is a genuine structured-bath memory-cost phenomenon.

The nonzero-$\mu$ extension sharpens this picture further. Allowing a finite chemical potential in the reset bath state does not generate an independent fixed-point particle-transport tradeoff; instead, it produces a filling-biased deformation of the same exact coherence-cost geometry. In particular, the retained coherence and the reset heat current remain controlled by the self-consistent occupation mismatch $P_{\mathrm{fp}}-n_F(\omega;\mu)$, so that their maxima are shifted systematically as $\mu$ is swept across the bath band. The inside-band case remains parametrically larger than the outside-band case, while the latter preserves the same qualitative structure on a compressed scale. Thus, a nonzero chemical potential acts as a genuine control knob for the fixed-point coherence-cost landscape, but it
does so by reshaping the original tradeoff rather than by opening a new thermodynamic channel.

Several extensions are immediate. One can replace the tight-binding bath by other spectral densities, incorporate static disorder, generalize the analysis to bosonic Gaussian baths, or include weak interactions within Hartree closures~\cite{Kumar2026Hartree}. One can also promote the heat-current analysis to a full Gaussian entropy-production analysis, drawing on the framework of coherence-to-work conversion developed in \cite{Korzekwa2016} to quantify the minimal thermodynamic cost of maintaining a reference coherence block in the bath. More broadly, the exact operating diagrams suggest a design principle for repeated-reset protocols: one should distinguish \emph{memory-optimal}, \emph{dissipation-optimal}, and \emph{memory-per-cost-optimal} reset channels rather than assuming that a single intervention protocol optimizes them all.

In conclusion, we have developed an exact theory of coherence-selective stroboscopic resetting in a quadratic open quantum system and solved it for the benchmark problem of a single fermionic level coupled to a semi-infinite tight-binding bath. The construction starts from the SPDM reset-map framework and the $\RI$/$\EC$ endpoints introduced in earlier work, but it asks a new question: how the exact fixed-point memory retained by a partially coherence-preserving reset channel is constrained by the thermodynamic price of resetting.

The answer is sharp. The fixed-point retained coherence $\CSEstar(\tau,\eta)$ is monotone in the coherence-memory parameter $\eta$, but the exact reset heat current $\JQstar(\tau,\eta)$ is generically nonmonotonic. The channel that stores the most coherence is therefore distinct from the channel that dissipates the most heat. Moreover, the exact operating diagrams show that the coherence-optimal and coherence-per-cost-optimal channels both lie near the coherence-preserving endpoint, while the heat-optimal channel is an interior point that drifts strongly with the reset interval. This establishes coherence-selective resetting as a distinct, exactly solvable, and practically interpretable control paradigm for structured-bath open quantum systems.

\section*{Data availability}

All data shown in the figures are available from the authors upon reasonable request.
\medskip

\begin{acknowledgments}
J.K. and T.A-N. have been supported by the Academy of Finland through its QTF Center of Excellence grant no. 312298 and by the European Union and the European Innovation Council through the Horizon Europe project QRC-4-ESP (grant No. 101129663), and EU Horizon Europe Quest project (No. 10116088). A.L. acknowledges support from the Leverhulme Trust Research Project Grant RPG-2025-063. 
\end{acknowledgments}

\appendix

\section{Exact derivation of the one-cycle affine map}
\label{app:map}

In this appendix we derive Eqs.~\eqref{eq:Pmap}-\eqref{eq:ymap} explicitly. Write the post-reset SPDM and one-cycle propagator as in Eqs.~\eqref{eq:rhon} and \eqref{eq:Ublock},
\begin{equation}
\rho_n = \begin{pmatrix}
P_n & \mathbf{x}_n\\
\mathbf{y}_n & C_0
\end{pmatrix};
\qquad
U = \begin{pmatrix}
u & \mathbf{v}\\
\mathbf{w} & X
\end{pmatrix}.
\end{equation}
First multiply $U\rho_n$:
\begin{equation}
U\rho_n=
\begin{pmatrix}
u P_n+\mathbf{v}\mathbf{y}_n & u\mathbf{x}_n+\mathbf{v}C_0\\
\mathbf{w}P_n+X\mathbf{y}_n & \mathbf{w}\mathbf{x}_n+XC_0
\end{pmatrix}.
\label{eq:Urho}
\end{equation}
Next multiply on the right by
\begin{equation}
U^{\dagger}=\begin{pmatrix}
u^{*} & \mathbf{w}^{\dagger}\\
\mathbf{v}^{\dagger} & X^{\dagger}
\end{pmatrix}.
\end{equation}
The top-left block of $\rho_n^{-}=U\rho_nU^{\dagger}$ is therefore
\begin{align}
P_n^{-} &= (u P_n+\mathbf{v}\mathbf{y}_n)u^{*} + (u\mathbf{x}_n+\mathbf{v}C_0)\mathbf{v}^{\dagger} \nonumber\\
&= |u|^2P_n + u^{*}\mathbf{v}\mathbf{y}_n + u\mathbf{x}_n\mathbf{v}^{\dagger} + \mathbf{v}C_0\mathbf{v}^{\dagger}.
\end{align}
Since the reset leaves the system block unchanged, $P_{n+1}=P_n^{-}$, which gives Eq.~\eqref{eq:Pmap}.

The top-right block is
\begin{align}
\mathbf{x}_n^{-} &= (u P_n+\mathbf{v}\mathbf{y}_n)\mathbf{w}^{\dagger} + (u\mathbf{x}_n+\mathbf{v}C_0)X^{\dagger}.
\end{align}
The reset multiplies this block by $\eta$, so
\begin{equation}
\mathbf{x}_{n+1}=\eta\mathbf{x}_n^{-}
\end{equation}
and Eq.~\eqref{eq:xmap} follows immediately.

Finally, the bottom-left block is
\begin{align}
\mathbf{y}_n^{-} &= \mathbf{w}(u^{*}P_n+\mathbf{x}_n\mathbf{v}^{\dagger}) + X(\mathbf{y}_nu^{*}+C_0\mathbf{v}^{\dagger}).
\end{align}
Again the reset multiplies this block by $\eta$, yielding Eq.~\eqref{eq:ymap}. The bottom-right block is the exact pre-reset bath SPDM,
\begin{align}
C_n^{-}
&= (\mathbf{w}P_n+X\mathbf{y}_n)\mathbf{w}^{\dagger} + (\mathbf{w}\mathbf{x}_n+XC_0)X^{\dagger} \nonumber\\
&= \mathbf{w}P_n\mathbf{w}^{\dagger} + X\mathbf{y}_n\mathbf{w}^{\dagger} + \mathbf{w}\mathbf{x}_nX^{\dagger}+XC_0X^{\dagger}.
\label{eq:Cnminus}
\end{align}
This is the formula used later for the heat current.

\section{\texorpdfstring{Leading small-$\tau$ asymptotics in explicit form}{Leading small-tau asymptotics in explicit form}}
\label{app:smalltau}

In this appendix we derive the small-$\tau$ formulas of the main text line by line, with no steps omitted. The starting point is the exact one-cycle update
\begin{equation}
\rho_n^{-}=U(\tau)\rho_n U^{\dagger}(\tau),
\end{equation}
with
\begin{equation}
U(\tau)=e^{-\ii M\tau}=I-\ii M\tau-\frac{1}{2}M^2\tau^2+\mathcal O(\tau^3).
\label{eq:Uexp}
\end{equation}
Substituting Eq.~\eqref{eq:Uexp} into $\rho_n^{-}=U\rho_nU^{\dagger}$ and keeping terms through first order gives
\begin{align}
\rho_n^{-}
&=(I-\ii M\tau)\rho_n(I+\ii M\tau)+\mathcal O(\tau^2)\nonumber\\
&=\rho_n-\ii M\rho_n\tau+\ii \rho_n M\tau+\mathcal O(\tau^2)\nonumber\\
&=\rho_n-\ii[M,\rho_n]\tau+\mathcal O(\tau^2).
\label{eq:rhofirstorder}
\end{align}
We now evaluate the matrix element $(0,k)$ in the bath eigenbasis of $M_{EE}$. In that basis the one-particle Hamiltonian has matrix elements
\begin{align}
M_{00}&=\omega_0;
\qquad
M_{kk'}=\omega_k\delta_{kk'};
\\
M_{0k}&=g_k;
\qquad
M_{k0}=g_k^*.
\end{align}
Immediately after a reset the SPDM takes the form
\begin{equation}
\rho_n=
\begin{pmatrix}
P_n & \mathbf{x}_n\\
\mathbf{y}_n & C_0
\end{pmatrix},
\end{equation}
with $C_0=\sum_k n_k |k\rangle\langle k|$ diagonal in the same basis. Therefore
\begin{equation}
(\rho_n)_{00}=P_n;
\quad
(\rho_n)_{0k}=x_{k,n};
\quad
(\rho_n)_{qk}=n_q\delta_{qk}.
\end{equation}
The first-order correction to $x_{k,n}$ is determined by
\begin{equation}
(\rho_n^{-})_{0k}=(\rho_n)_{0k}-\ii [M,\rho_n]_{0k}\tau+\mathcal O(\tau^2).
\label{eq:x0kfromcomm}
\end{equation}
Now
\begin{equation}
[M,\rho_n]_{0k}=\sum_{\gamma}M_{0\gamma}(\rho_n)_{\gamma k}-\sum_{\gamma}(\rho_n)_{0\gamma}M_{\gamma k}.
\label{eq:comm00kdef}
\end{equation}
We evaluate the two sums separately.

For the first sum,
\begin{align}
\sum_{\gamma}M_{0\gamma}(\rho_n)_{\gamma k}
&=M_{00}(\rho_n)_{0k}+\sum_q M_{0q}(\rho_n)_{qk}\nonumber\\
&=\omega_0 x_{k,n}+\sum_q g_q n_q\delta_{qk}\nonumber\\
&=\omega_0 x_{k,n}+g_k n_k.
\label{eq:firstsumxk}
\end{align}
For the second sum,
\begin{align}
\sum_{\gamma}(\rho_n)_{0\gamma}M_{\gamma k}
&=(\rho_n)_{00}M_{0k}+\sum_q(\rho_n)_{0q}M_{qk}\nonumber\\
&=P_n g_k+\sum_q x_{q,n}\omega_q\delta_{qk}\nonumber\\
&=P_n g_k+\omega_k x_{k,n}.
\label{eq:secondsumxk}
\end{align}
Substituting Eqs.~\eqref{eq:firstsumxk} and \eqref{eq:secondsumxk} into Eq.~\eqref{eq:comm00kdef} gives
\begin{align}
[M,\rho_n]_{0k}
&=(\omega_0-\omega_k)x_{k,n}+g_k(n_k-P_n).
\label{eq:commxkexplicit}
\end{align}
Insert this into Eq.~\eqref{eq:x0kfromcomm}:
\begin{align}
x_k^{-}
&=x_{k,n}-\ii\big[(\omega_0-\omega_k)x_{k,n}+g_k(n_k-P_n)\big]\tau+\mathcal O(\tau^2)\nonumber\\
&=x_{k,n}-\ii(\omega_0-\omega_k)x_{k,n}\tau+\ii g_k(P_n-n_k)\tau+\mathcal O(\tau^2).
\label{eq:xkminusstep}
\end{align}
This is the pre-reset coherence to first order. The reset multiplies the $SE$ block by $\eta$, so
\begin{equation}
x_{k,n+1}=\eta x_k^{-}.
\end{equation}
Using Eq.~\eqref{eq:xkminusstep},
\begin{align}
x_{k,n+1}
&=\eta x_{k,n}-\ii\eta(\omega_0-\omega_k)x_{k,n}\tau \nonumber\\&+\ii\eta g_k(P_n-n_k)\tau+\mathcal O(\tau^2),
\end{align}
which is Eq.~\eqref{eq:xkrec} of the main text.

We now derive the fixed-point expression. At a stroboscopic fixed point,
\begin{equation}
x_{k,n+1}=x_{k,n}=x_k^{\fp};
\qquad
P_n=P_{\fp}.
\end{equation}
Therefore
\begin{equation}
x_k^{\fp}=\eta x_k^{\fp}-\ii\eta(\omega_0-\omega_k)x_k^{\fp}\tau+\ii\eta g_k(P_{\fp}-n_k)\tau+\mathcal O(\tau^2).
\end{equation}
Bring the terms containing $x_k^{\fp}$ to the left-hand side:
\begin{align}
x_k^{\fp}-\eta x_k^{\fp}+\ii\eta(\omega_0-\omega_k)x_k^{\fp}\tau
&=\ii\eta g_k(P_{\fp}-n_k)\tau+\mathcal O(\tau^2).
\end{align}
Factor out $x_k^{\fp}$:
\begin{equation}
\Big[(1-\eta)+\ii\eta(\omega_0-\omega_k)\tau\Big]x_k^{\fp}=\ii\eta g_k(P_{\fp}-n_k)\tau+\mathcal O(\tau^2).
\label{eq:prexkfp}
\end{equation}
Dividing both sides by the bracket yields
\begin{equation}
x_k^{\fp}(\tau,\eta)=\frac{\ii\eta g_k(P_{\fp}-n_k)\tau}{(1-\eta)+\ii\eta(\omega_0-\omega_k)\tau}+\mathcal O(\tau^2),
\end{equation}
which is Eq.~\eqref{eq:xkfp}. To obtain the modulus squared, use
\begin{align}
\left|\frac{a}{b}\right|^2&=\frac{|a|^2}{|b|^2};
\\
|(1-\eta)+\ii\eta(\omega_0-\omega_k)\tau|^2&=(1-\eta)^2+\eta^2(\omega_0-\omega_k)^2\tau^2.
\end{align}
Hence
\begin{align}
|x_k^{\fp}(\tau,\eta)|^2
&=\frac{\eta^2|g_k|^2(P_{\fp}-n_k)^2\tau^2}{(1-\eta)^2+\eta^2(\omega_0-\omega_k)^2\tau^2}+\mathcal O(\tau^3),
\end{align}
which is Eq.~\eqref{eq:xkfpmod}.

The same calculation also gives the small-$\tau$ scaling of the total retained coherence. Summing the previous expression over $k$ gives
\begin{align}
\CSE(\tau,\eta)&=\sum_k |x_k^{\fp}(\tau,\eta)|^2 \nonumber\\
&=\tau^2\sum_k\frac{\eta^2|g_k|^2(P_{\fp}-n_k)^2}{(1-\eta)^2+\eta^2(\omega_0-\omega_k)^2\tau^2}+\mathcal O(\tau^3).
\label{eq:CSEsmalltauappendix}
\end{align}
For fixed $\eta<1$ and $\tau\to 0$, the denominator is dominated by $(1-\eta)^2$, so
\begin{equation}
\CSE(\tau,\eta)\sim \frac{\eta^2\tau^2}{(1-\eta)^2}\sum_k |g_k|^2(P_{\fp}-n_k)^2.
\label{eq:CSEsmalltauscaling}
\end{equation}
Thus, at fixed $\eta<1$, the retained coherence vanishes quadratically in the reset interval, while its prefactor diverges as $\eta\to 1^{-}$. This is the asymptotic statement behind the strong upturn seen in the exact numerics.

\section{Derivation of retained coherence, reset heat current, and coherence efficiency}
\label{app:observables}

In this appendix we derive the main-text observables introduced in Sec.~\ref{sec:tb}. The goal is to show explicitly how Eqs.~\eqref{eq:CSEdef}--\eqref{eq:Rdef} arise from the exact fixed-point SPDM.

\subsection{Retained coherence and coherence spectrum}

The exact post-reset fixed-point SPDM has the block form
\begin{equation}
\rho_{\fp}=\begin{pmatrix}
P_{\fp} & \mathbf{x}_{\fp}\\
\mathbf{y}_{\fp} & C_0
\end{pmatrix},
\end{equation}
where the row vector $\mathbf{x}_{\fp}$ contains all system--environment coherences after the reset. In the bath eigenbasis of $M_{EE}$, its components are
\begin{equation}
x_k^{\fp}=\langle 0|\rho_{\fp}|k\rangle.
\end{equation}
Because each mode carries a distinct coherence amplitude, the most microscopic quantity is the mode-resolved spectrum
\begin{equation}
\Sc(\omega_k;\tau,\eta)=|x_k^{\fp}(\tau,\eta)|^2.
\end{equation}
Passing to a continuum bath, one writes $k\mapsto \omega$ and uses the system--bath spectral density $J(\omega)$, so the continuum guide quantity is
\begin{equation}
\Sc(\omega;\tau,\eta)\approx \frac{\eta^2\tau^2 J(\omega)[P_{\fp}-n_F(\omega)]^2}{(1-\eta)^2+\eta^2(\omega_0-\omega)^2\tau^2},
\end{equation}
which is Eq.~\eqref{eq:Scomeguide}. The total retained coherence is then obtained by summing or integrating the spectrum:
\begin{equation}
\CSE(\tau,\eta)=\sum_k |x_k^{\fp}(\tau,\eta)|^2,
\end{equation}
or, in continuum language,
\begin{equation}
\CSE(\tau,\eta)=\int d\omega\,\Sc(\omega;\tau,\eta).
\end{equation}
This is Eq.~\eqref{eq:CSEdef} of the main text. It is simply the quadratic norm of the fixed-point $SE$ coherence block.

\subsection{Exact pre-reset bath block}

To derive the heat current, we first need the exact pre-reset bath SPDM. Starting from the exact fixed-point post-reset state $\rho_{\fp}$, one cycle of unitary evolution gives
\begin{equation}
\rho_{\fp}^{-}=U(\tau)\rho_{\fp}U^{\dagger}(\tau).
\end{equation}
From the block multiplication derived in Appendix~\ref{app:map}, the bottom-right block is
\begin{equation}
C_{\fp}^{-}=\mathbf{w}P_{\fp}\mathbf{w}^{\dagger}+X\mathbf{y}_{\fp}\mathbf{w}^{\dagger}+\mathbf{w}\mathbf{x}_{\fp}X^{\dagger}+XC_0X^{\dagger}.
\end{equation}
Each term has a clear meaning:
\begin{enumerate}
\item $\mathbf{w}P_{\fp}\mathbf{w}^{\dagger}$ is the bath population generated directly from the occupied system level during the unitary segment;
\item $X\mathbf{y}_{\fp}\mathbf{w}^{\dagger}$ and $\mathbf{w}\mathbf{x}_{\fp}X^{\dagger}$ are the two terms linear in the retained coherence;
\item $XC_0X^{\dagger}$ is the dressed propagation of the reference bath block through the unitary segment.
\end{enumerate}
Thus the exact pre-reset bath SPDM depends not only on the system occupation but also explicitly on the retained coherence block. This is the microscopic origin of the coherence--cost tradeoff.

\subsection{Reset heat and heat current}

The reset operation restores the bath block from $C_{\fp}^{-}$ to $C_0$. The energy removed from the bath by that reset step is therefore
\begin{equation}
\Qsupstar(\tau,\eta)=\Tr_E\big[M_{EE}(C_{\fp}^{-}-C_0)\big],
\end{equation}
which is Eq.~\eqref{eq:Qreset}. The corresponding heat current per cycle is obtained by dividing by the cycle length,
\begin{equation}
\JQstar(\tau,\eta)=\frac{\Qsupstar(\tau,\eta)}{\tau}.
\end{equation}
This is Eq.~\eqref{eq:JQdef}. Because $C_{\fp}^{-}$ contains terms linear in $\mathbf{x}_{\fp}$ and $\mathbf{y}_{\fp}$, the reset heat is not a function of $P_{\fp}$ alone. It responds directly to the retained system--environment coherence.

To see this at leading order, evaluate the coherence-dependent contribution to $C_{\fp}^{-}$ in the bath eigenbasis. The two terms linear in coherence are
\begin{equation}
X\mathbf{y}_{\fp}\mathbf{w}^{\dagger}+\mathbf{w}\mathbf{x}_{\fp}X^{\dagger}.
\end{equation}
Using $X=I+\mathcal O(\tau)$ and $\mathbf{w}_k=-\ii g_k^*\tau+\mathcal O(\tau^2)$, one finds
\begin{equation}
Q_{\mathrm{coh}}(\tau,\eta)=-\ii\tau\sum_k \omega_k\big(g_k^*x_k^{\fp}-x_k^{\fp*}g_k\big)+\mathcal O(\tau^2).
\label{eq:Qcohappendix}
\end{equation}
We now insert Eq.~\eqref{eq:xkfp}. First,
\begin{equation}
g_k^*x_k^{\fp}=\frac{\ii\eta |g_k|^2(P_{\fp}-n_k)\tau}{(1-\eta)+\ii\eta(\omega_0-\omega_k)\tau}.
\end{equation}
Its complex conjugate is
\begin{equation}
x_k^{\fp*}g_k=\frac{-\ii\eta |g_k|^2(P_{\fp}-n_k)\tau}{(1-\eta)-\ii\eta(\omega_0-\omega_k)\tau}.
\end{equation}
Therefore
\begin{align}
g_k^*x_k^{\fp}-x_k^{\fp*}g_k
&=\ii\eta |g_k|^2(P_{\fp}-n_k)\tau\bigg[\frac{1}{(1-\eta)+\ii\eta\Delta_k\tau} \nonumber\\ &+\frac{1}{(1-\eta)-\ii\eta\Delta_k\tau}\bigg]\\
&=\ii\eta |g_k|^2(P_{\fp}-n_k)\tau\frac{2(1-\eta)}{(1-\eta)^2+\eta^2\Delta_k^2\tau^2},
\end{align}
where $\Delta_k=\omega_0-\omega_k$. Substituting into Eq.~\eqref{eq:Qcohappendix} gives
\begin{equation}
Q_{\mathrm{coh}}(\tau,\eta)=\tau^2\sum_k \omega_k\frac{2\eta(1-\eta)|g_k|^2(P_{\fp}-n_k)}{(1-\eta)^2+\eta^2(\omega_0-\omega_k)^2\tau^2}+\mathcal O(\tau^2_{\mathrm{sub}}).
\label{eq:Qcohfinalappendix}
\end{equation}
This is the explicit leading-order coherence-induced contribution to the reset heat. It is the derivation behind the statements in Sec.~\ref{sec:tb} that retained coherence enters the thermodynamic cost directly.

\begin{figure*}[t]
\centering
\includegraphics[width=0.48\textwidth]{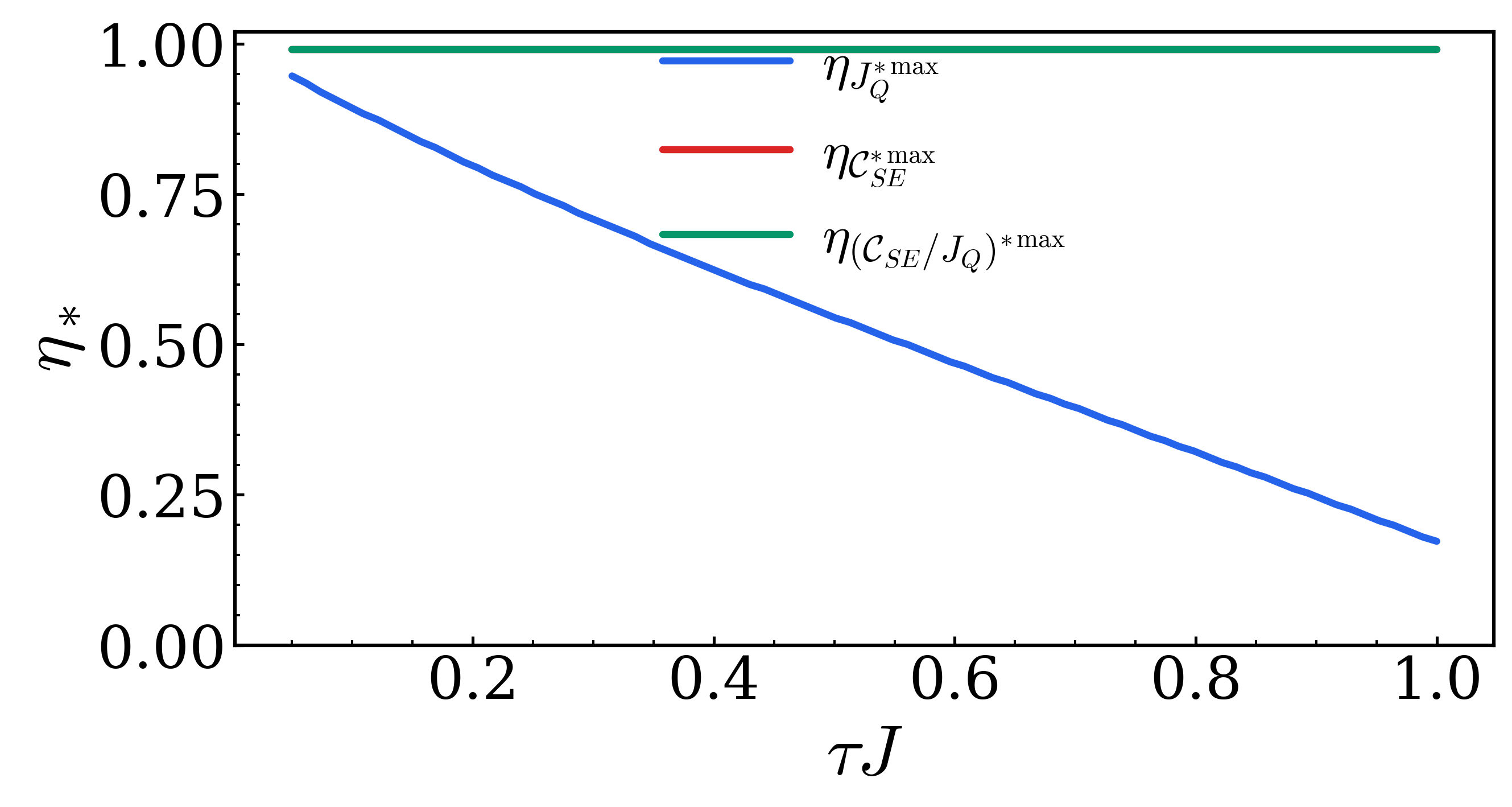}\hfill
\includegraphics[width=0.48\textwidth]{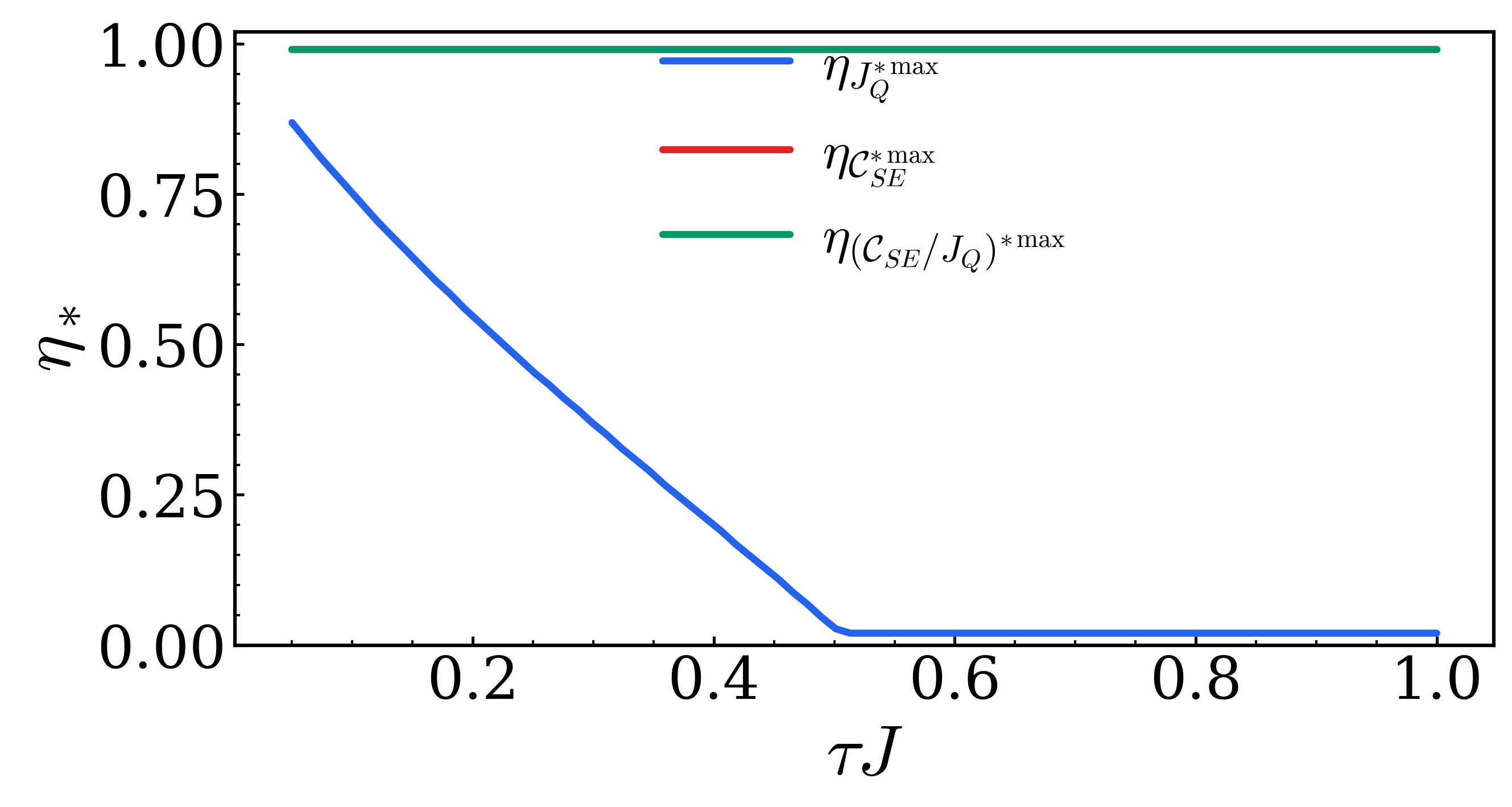}\\[1ex]
\includegraphics[width=0.48\textwidth]{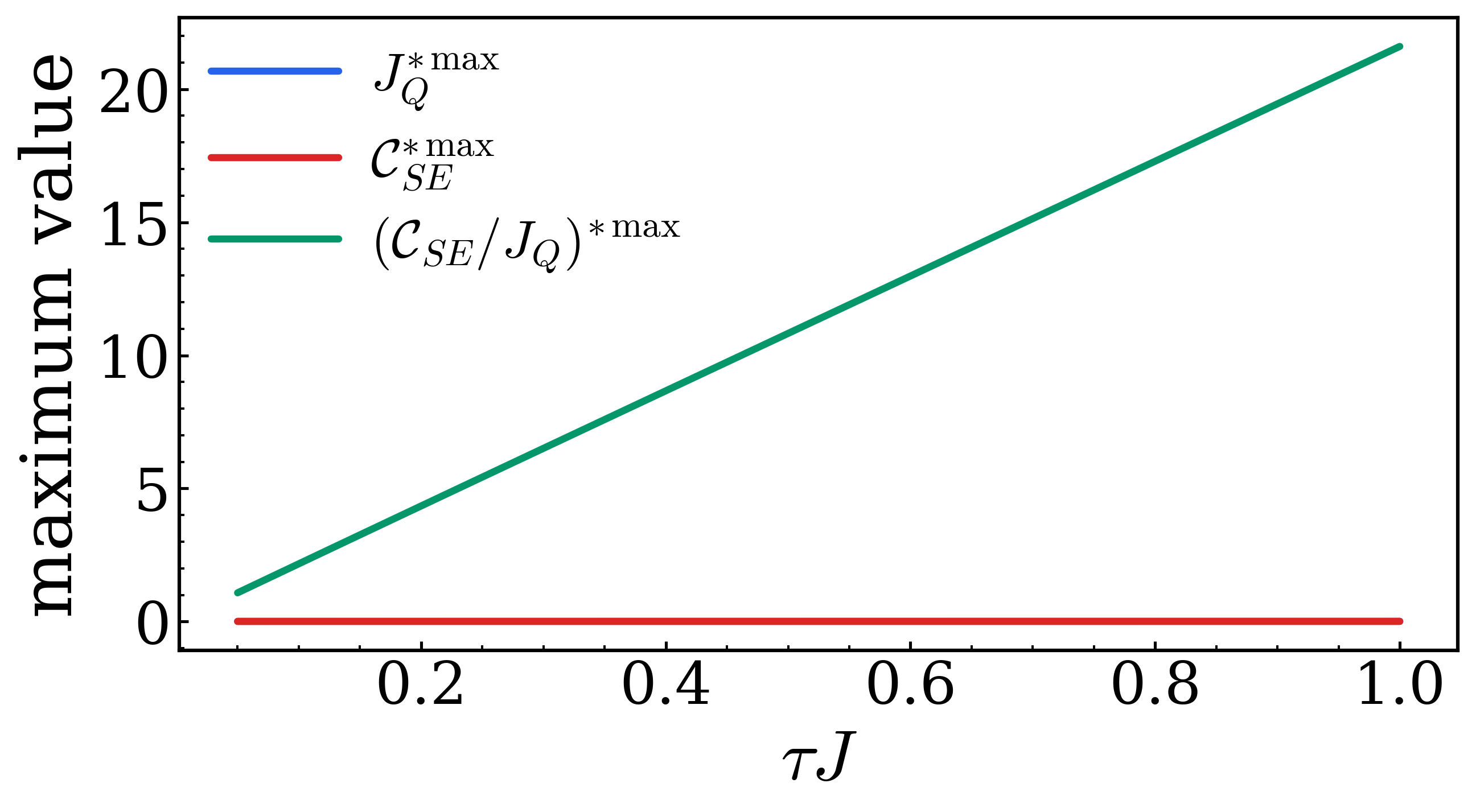}\hfill
\includegraphics[width=0.48\textwidth]{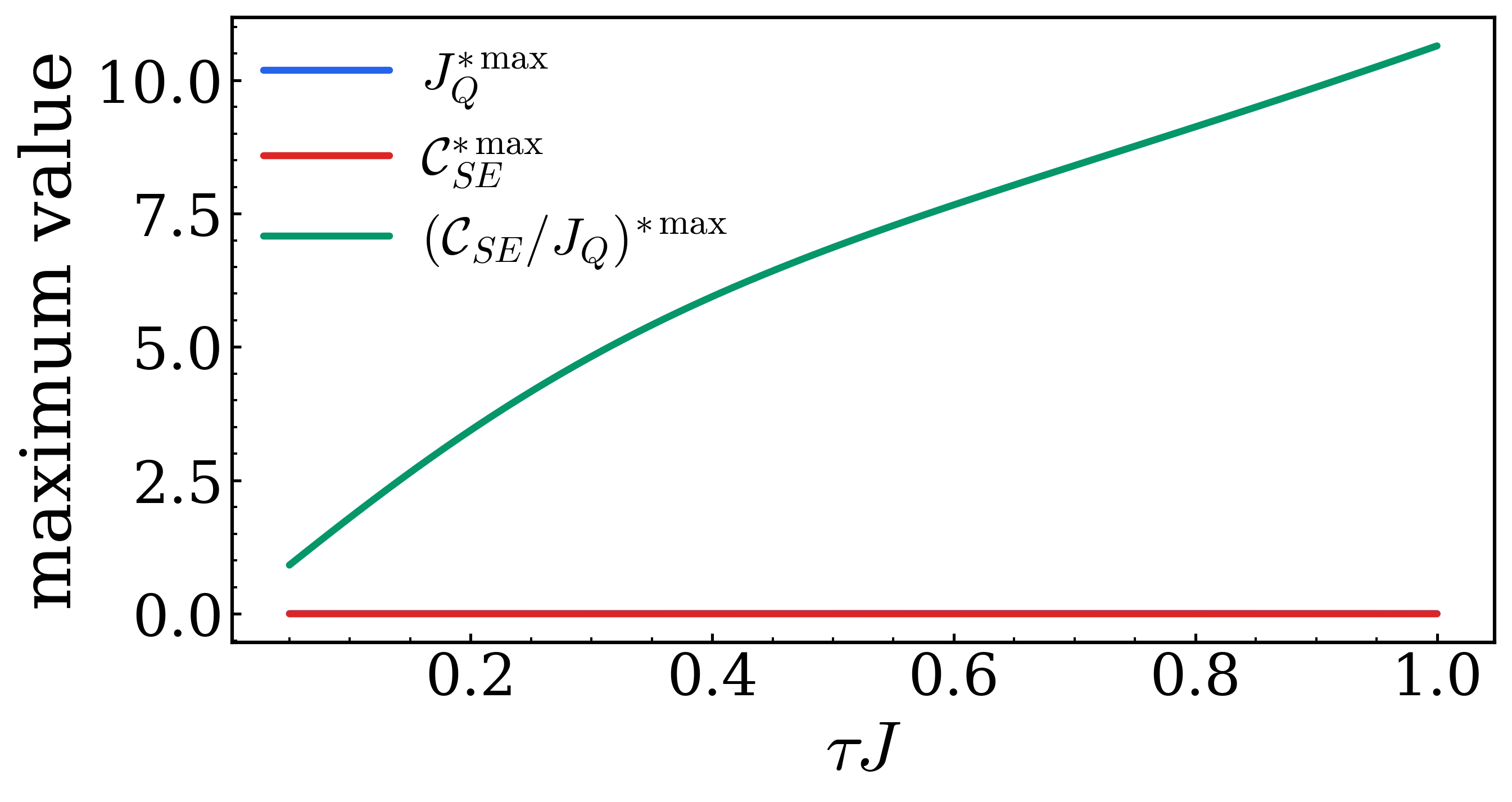}
\caption{Exact operating-point extraction. Top row: maximizing values of $\eta$ for the heat current, retained coherence, and coherence-efficiency ratio. Bottom row: the corresponding maximal values. Left column: inside-band level $\omega_0/J=0.8$. Right column: outside-band level $\omega_0/J=3.0$. In both cases the coherence-optimal and coherence-per-cost-optimal channels stay near the coherence-preserving endpoint, whereas the heat-optimal channel drifts strongly with $\tau$ toward smaller $\eta$.}
\label{fig:operating}
\end{figure*}
\subsection{Coherence efficiency and operating points}

The coherence-efficiency ratio is defined by
\begin{equation}
R^{*}(\tau,\eta)=\frac{\CSEstar(\tau,\eta)}{\JQstar(\tau,\eta)}.
\end{equation}
This quantity has no independent microscopic derivation: it is a composite operational figure of merit built from the exact retained coherence and exact reset heat current. Its role is to quantify how much retained coherence is obtained per unit reset-heat cost.

For each fixed $\tau$, the operating points
\begin{equation}
\eta_{\JQ}^{\max}(\tau);\qquad \eta_{\CSE}^{\max}(\tau);\qquad \eta_R^{\max}(\tau)
\end{equation}
are defined by the optimization conditions
\begin{align}
\JQ(\tau,\eta_{\JQ}^{\max})&=\max_{0\le \eta\le 1}\JQ(\tau,\eta);\\
\CSE(\tau,\eta_{\CSE}^{\max})&=\max_{0\le \eta\le 1}\CSE(\tau,\eta);\\
R(\tau,\eta_R^{\max})&=\max_{0\le \eta\le 1}R(\tau,\eta).
\end{align}
These operating points are determined numerically from the exact fixed-point map. The resulting exact tradeoff curves in the main text show that the three optima do not coincide: the channel that retains the most coherence is generically not the channel that maximizes the reset heat, and the channel that optimizes coherence per unit cost lies near, but not necessarily exactly at, the full-coherence endpoint.

Figure~\ref{fig:operating} shows the corresponding maximizing values of $\eta$. For the inside-band level, the heat-optimal channel moves steadily to smaller $\eta$ as $\tau$ increases, while both the coherence-optimal and coherence-per-cost-optimal channels remain essentially pinned near the coherence-preserving endpoint. Outside the band the same trend persists, and the heat-optimal $\eta$ collapses even more rapidly toward smaller values as $\tau$ increases. This exact operating-point extraction condenses the central message of the paper into a simple design rule: partial coherence preservation is thermodynamically most expensive only at an interior operating point, whereas near-complete coherence preservation is best for both memory storage and memory-per-cost.

\section{Semi-infinite tight-binding spectral density}

\label{app:tb}

We derive Eq.~\eqref{eq:Jomega}. For the semi-infinite chain, the normalized eigenmodes are Eq.~\eqref{eq:standing} and the dispersion is Eq.~\eqref{eq:dispersion}. The coupling to the system level is $g(k)=t_{\rm c}\sqrt{2/\pi}\sin k$. The spectral density is defined by
\begin{equation}
J(\omega)=\int_0^{\pi} \dd k\, |g(k)|^2\,\delta\big(\omega-\omega(k)\big).
\end{equation}
Substituting the explicit expressions gives
\begin{equation}
J(\omega)=\frac{2t_{\rm c}^2}{\pi}\int_0^{\pi} \dd k\,\sin^2 k\,\delta(\omega+2J\cos k).
\end{equation}
Let $f(k)=\omega+2J\cos k$. The root condition $f(k_0)=0$ implies
\begin{equation}
\cos k_0 = -\frac{\omega}{2J},
\end{equation}
which has a solution only when $|\omega|<2J$. Using the identity
\begin{equation}
\delta\big(f(k)\big)=\sum_{k_0}\frac{\delta(k-k_0)}{|f'(k_0)|},
\end{equation}
with $f'(k)=-2J\sin k$, we obtain
\begin{equation}
J(\omega)=\frac{2t_{\rm c}^2}{\pi}\frac{\sin^2 k_0}{2J|\sin k_0|} = \frac{t_{\rm c}^2}{\pi J}|\sin k_0|.
\end{equation}
Now
\begin{equation}
\sin^2 k_0 = 1-\cos^2 k_0 = 1-\frac{\omega^2}{4J^2}=\frac{4J^2-\omega^2}{4J^2}.
\end{equation}
Hence
\begin{equation}
|\sin k_0|=\frac{\sqrt{4J^2-\omega^2}}{2J},
\end{equation}
and therefore
\begin{equation}
J(\omega)=\frac{t_{\rm c}^2}{2\pi J^2}\sqrt{4J^2-\omega^2};\qquad |\omega|<2J,
\end{equation}
while $J(\omega)=0$ outside the band.

\section{Numerical implementation}
\label{app:numerics}

The semi-infinite chain is represented numerically by a finite open chain of $N_{\rm b}$ bath sites. For each value of $(\tau,\eta,\omega_0)$ we build the one-particle Hamiltonian matrix of dimension $1+N_{\rm b}$, compute the exact propagator $U(\tau)=e^{-\ii M\tau}$, assemble the affine map of Eqs.~\eqref{eq:Pmap}-\eqref{eq:ymap}, and solve the linear fixed-point equation Eq.~\eqref{eq:Wfp}. The exact observables $\CSE$, $\JQ$, $P_{\fp}$, and $R$ are then evaluated from the fixed point. The spectra and line-cut figures use a larger truncation, while the two-parameter heat maps use a slightly smaller but still converged truncation to keep the full parameter sweep economical. In practice we found that the qualitative operating-point structure is stable once the chain is long enough that the one-cycle propagator does not probe the far boundary on the timescales shown. This is the standard convergence criterion for finite representations of semi-infinite tight-binding baths. The numerical figures included here are therefore exact for the truncated model and converged with respect to that truncation at the level relevant for the operating diagrams reported in the main text.

\bibliography{coherence_selective}

\end{document}